\documentclass[structabstract]{aa}  
\usepackage{natbib}
\usepackage{graphicx}
\usepackage{txfonts}
\newcommand{\dg}{$^\circ$}

\newcommand{\hr}{HR~5999}
\newcommand{\m}{$\mu$m}
\newcommand{\uv}{\textit{UV}~}
\newcommand{\Mira}{\textit{MiRA}}
\newcommand{\mathe}{\mathrm{e}}
\newcommand{\mathi}{\mathrm{i}}
\begin{document}
   \title{A low optical depth region in the inner disk of the Herbig~Ae
star HR~5999\thanks{Based on observations collected at the VLTI
   (ESO   Paranal,  Chile)   with  programs   080.C-0056,  083.C-0298,
   083.C-0144,   083.C-0334,   083.C-0170,   083.C-0857,   083.C-0864,
   083.C-0602, 084.C-0590, 085.C-0769, 085.C-0502}}

   \author{M.~Benisty\inst{1,2},   S.~Renard\inst{3},  A.~Natta\inst{1},
   J.P.~Berger\inst{4,3},     F.~Massi\inst{1},     F.~Malbet\inst{3},
   P.J.V.~Garcia\inst{5},      A.~Isella\inst{6},     A.~M\'erand\inst{4},
   J.L.~Monin\inst{3},    L.~Testi\inst{1,7}, E.~Thi\'ebaut\inst{8},   M.~Vannier\inst{9},
   G.~Weigelt\inst{10}} 

\institute{INAF-Osservatorio Astrofisico di Arcetri, Largo E.~Fermi
          5, 50125 Firenze, Italy
          \and
          Max Planck Institut f\"ur Astronomie, Konigst\"uhl 17, 69117
          Heidelberg, Germany - \email{benisty@mpia.de}
          \and
          Laboratoire d'Astrophysique de  Grenoble, CNRS-UJF UMR 5571,
          414 rue de la Piscine, 38400 St Martin d'H\`eres, France         
          \and
          European  Southern Observatory, Casilla  19001, Santiago
          19, Chile
          \and
          Universidade do Porto,  Faculdade de Engenharia, SIM Unidade
          FCT 4006, Rua Dr. Roberto Frias, s/n P-4200-465 Porto, Portugal
          \and
          Caltech, MC 249-17, 1200  East California Blvd, Pasadena, CA
          91125, USA
          \and
           European  Southern Observatory,	Karl-Schwarzschild-Strasse 2, 85748 Garching, Germany
          \and
          Centre  de Recherche  Astrophysique de  Lyon, CNRS-UCBL-ENSL
          UMR5574, 69561 St Genis Laval, France
          \and
          Laboratoire A.~H.~Fizeau, UMR~6525, Universit\'e de Nice-Sophia
          Antipolis, Parc Valrose, 06108 Nice Cedex 02, France
          \and
          Max Planck  Institut f\"ur Radioastronomie,  Auf dem H\"ugel
          69, 53121 Bonn, Germany\\ }
        
\date{Received 8 November 2010 / Accepted 8 April 2011}

% \abstract{}{}{}{}{} 
% 5 {} token are mandatory
 
  \abstract
  % context heading (optional)
  % {} leave it empty if necessary  
  {Circumstellar  disks  surrounding young  stars  are  known  to be  the
    birthplaces of  planetary systems, and  the innermost astronomical
    unit   is    of   particular   interest.      Near-infrared
    interferometric studies have revealed a complex morphology for the
    close environment surrounding Herbig~Ae stars. }
                                % aims heading (mandatory)
   {We present new long-baseline spectro-interferometric observations
     of the  Herbig~Ae star,  HR~5999, obtained in  the $H$ and  $K$ bands
     with  the  AMBER instrument  at  the  VLTI,  and aim  to  produce
     near-infrared images at the sub-AU spatial scale. } 
% methods heading (mandatory)
   {We spatially resolve the circumstellar material and reconstruct images in
 the $H$ and $K$ bands using the MiRA algorithm. In addition, we interpret the
     interferometric observations using models that assume that the near-infrared
 excess  is dominated  by the  emission  of a  circumstellar disk.  We
     compare the  images reconstructed  from the VLTI  measurements to
 images obtained using simulated model data. }
  % results heading (mandatory)
   {The $K$-band image reveals three main elements: a ring-like feature located
   at $\sim$0.65~AU, a low surface brightness region inside 0.65~AU,
   and a central  spot.  At the maximum angular resolution of our observations  ($B/\lambda \sim$1.3~mas), the ring is resolved while the central spot is only marginally resolved, preventing us from revealing the exact morphology of the circumstellar environment.  We suggest that  the ring  traces 
   silicate condensation, i.e., an opacity change, in a circumstellar disk around
 HR~5999.    We build a model that includes a ring at the silicate sublimation radius and an inner disk of low surface brightness responsible for a large amount of the near-infrared continuum emission. The model successfully fits the SED, visibilities, and closure phases in the $H$ and  $K$ bands, and provides evidence of a low surface brightness region inside the silicate sublimation radius.}  
  % conclusions heading (optional), leave it empty if necessary 
 {This study provides milli-arcsecond resolution images of the environment of
   HR~5999 and additional evidence  that in Herbig~Ae~stars, there is
   material in a low surface brightness region, probably a low optical depth region, located inside the silicate
   sublimation radius and of  unknown nature. The possibility that the
   formation of such a region in a thick disk is related
   to disk evolution should  be investigated. }
 
   \keywords{Stars:  individual:  \hr  - circumstellar matter - Techniques: interferometric}

   \authorrunning{Benisty et al.}
   \titlerunning{HR~5999 inner disk}

   \maketitle
%
%________________________________________________________________

\section{Introduction}
Circumstellar  disks  surrounding young  stars  are  known  to be  the
birthplaces of planetary systems.   To understand planet formation, it
is crucial to study the physical conditions in the disks in which they
form, and the processes that control  the evolution of gas and dust.  The
innermost astronomical unit (AU) is of particular interest because 
star-disk interactions, such as mass accretion and
ejection, occur there.  In the past decade,
near-infrared  (NIR) long-baseline interferometry  has allowed  us to
directly probe  matter within the first AU.  The first interferometric
studies  of  intermediate-mass  young  stars,  the  Herbig~AeBe  stars
(HAeBe), have shown that the characteristic sizes of the emission in the NIR were
larger than 
expected with  classical accretion disk  models \citep{millangabet01},
and   were  found  to  be correlated  with the   stellar  luminosity
\citep{monnier02}.  This supports the idea that the NIR emission is 
dominated  by   the thermal  emission  of  hot  dust   heated  by  stellar
radiation. \cite{natta01} suggested that an inner, optically thin
cavity due  to dust sublimation exists  in the disk. At  the edge of
this region, where dust condensates, the disk is expected to puff up
because of   direct      illumination      from     the      star
\citep{dullemond01,isella05,tannirkulam07,kama09}, explaining the size-luminosity law
derived for Herbig~Ae (and late Be) stars by \cite{monnier02}.   
On  the basis of a  small  number of  interferometric  observations,  simple
geometrical models  were proposed to explain the  global morphology of
these    regions   \citep[e.g.,][]{eisner04,monnier05}.

However, as higher quality data sets have become available, it has become clear
that the  regions probed by  NIR interferometry are much  more complex
and  that   a  deeper   understanding  requires  the   combination  of
photometric and multi-wavelength interferometric measurements together
with more sophisticated models.   Using the very long CHARA baselines,
\cite{tannirkulam} found that the $K$-band observations at  the milli-arcsecond  (mas)
resolution of the Herbig~Ae star HD~163296 could not be reproduced using models
where the majority of the $K$-band emission arises in a dust rim, but
that  an additional NIR  emission inside  the dust  sublimation radius
was needed to explain the visibilities and the SED.  They interpreted this
additional emission as  being produced by gas, as  suggested for other
Herbig~AeBe  stars when using  spectro-interferometry \citep[e.g.,][]{eisner07,isella08,kraus08}.     Using    the  largest
interferometric   dataset  so   far,  \cite{benisty10}   suggested  an
alternative scenario  for HD~163296 of  an optically thin  inner disk,
made of  refractory grains  such as  iron up to  0.5~AU, as  the major
source of the NIR emission in HD~163296. 

With the advent  of interferometers with more than  two telescopes and
of significantly high spectral and spatial resolution (e.g., 
VLTI/AMBER, CHARA/MIRC), the  number of interferometric observations
has significantly increased allowing the first images to be
reconstructed with aperture synthesis techniques.  Images of stellar surfaces
\citep[e.g.,][]{haubois09},  binaries  \citep[e.g.,][]{zhao08},  and
circumstellar        shells        around       evolved        stars
\citep[e.g.,][]{leBouquin09} have been obtained. Two
circumstellar disks have been imaged, around an intermediate-mass
young  star  \citep{renard10a}  and  around  a  massive  young  star
\citep{kraus10}. 

\begin{figure*}[t]
 \centering
\begin{tabular}{cc}
  \includegraphics[width=0.4\textwidth]{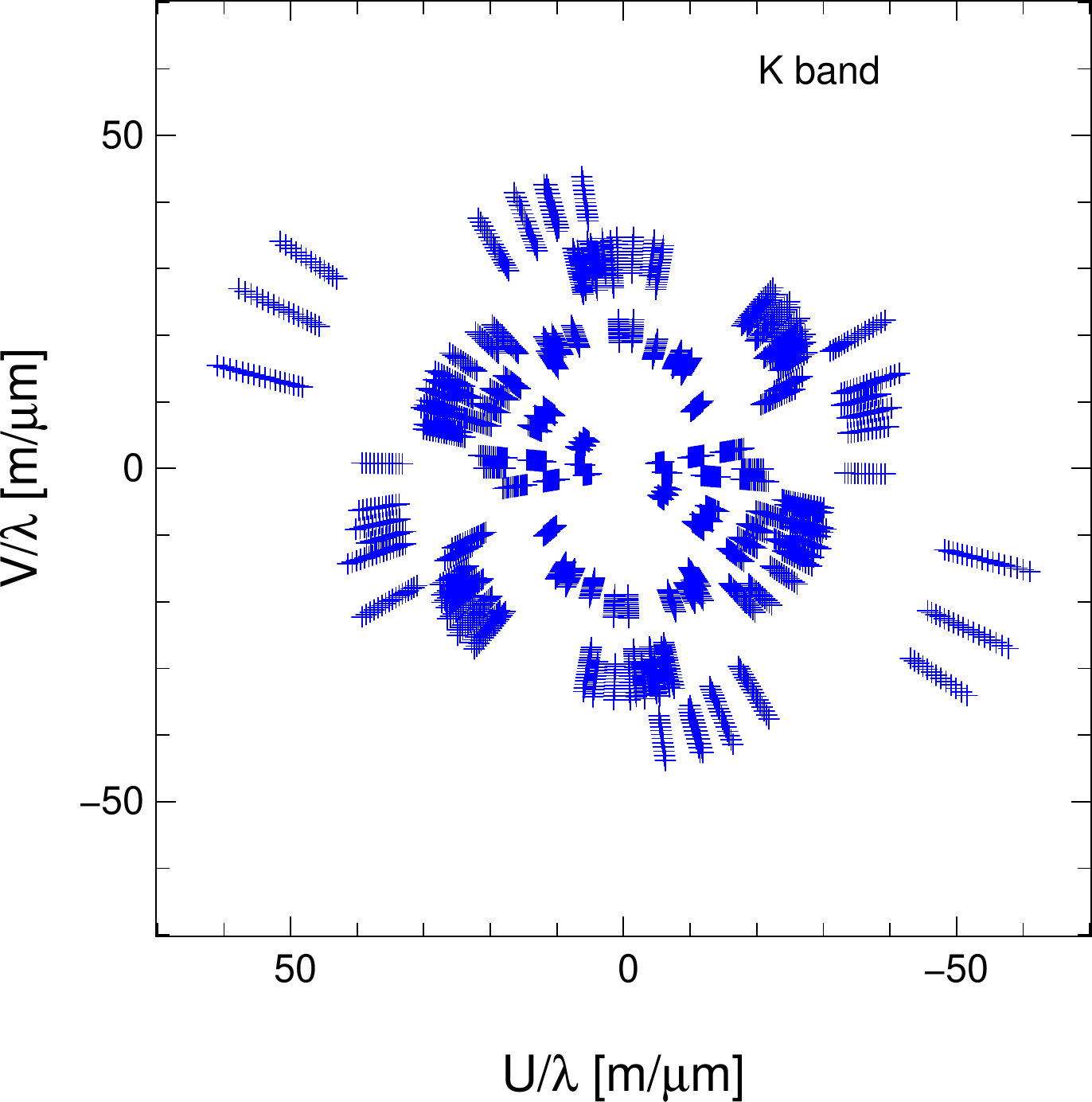} & 
  \includegraphics[width=0.4\textwidth]{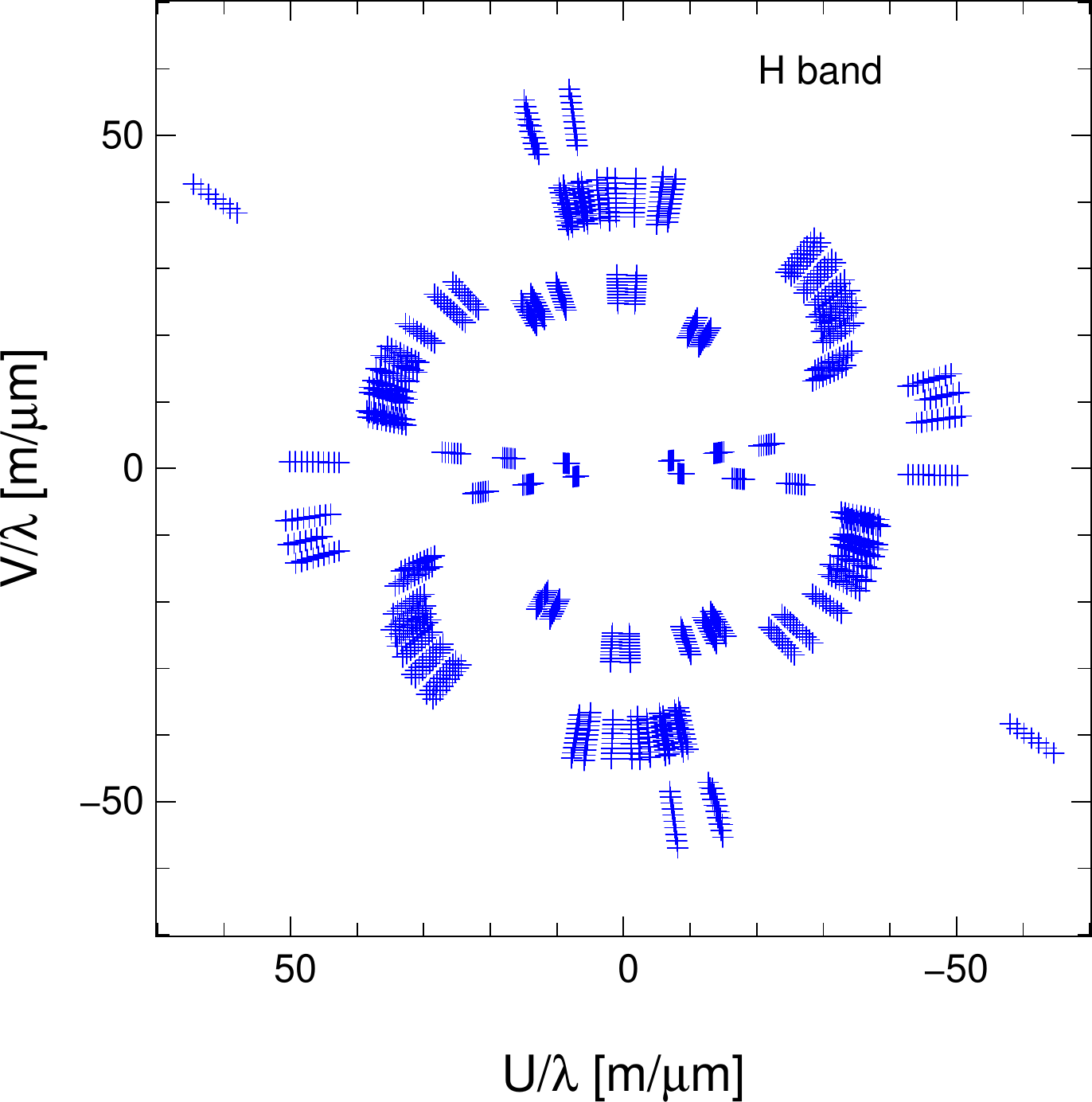}
\end{tabular}
  \caption{\label{fig:uv} \uv coverage obtained in the $K$-band (left) and in the $H$-band (right).} 
\end{figure*}

In this  study, we  present NIR  images and an  analysis of  the inner
environment surrounding 
the Herbig~Ae star \hr~(V856~Sco, HD144668). The star lies in the Lupus
star formation complex, at  a distance of 208$^{+50}_{-30}$~pc.  It is
part  of  a quadruple  system,   its  closest detected  companion
(Rossiter 3930;  $\Delta  V, H, K$ = 4.6, 3.1, 3.6 mag \citep{stecklum95}) being at 1.4''.  \hr~was identified as a 3-4~M$_{\odot}$
star, that is $\sim$0.5~Myr  old,  with  a~A7  III-IV spectral  type
\citep{tjin89},         T$_{\rm{eff}}$=7925~K,        L=87~L$_{\odot}$
\citep{vanboekel05}, and A$_{\rm{v}}$=0.49~mag. Photometric and spectroscopic variability was observed in the optical 
\citep{perez92} and UV \citep{perez93} and interpreted as evidence of 
accreting  gas  in  a  boundary  layer  in  a  highly  inclined  disk.
\cite{hubrig07}  detected a  magnetic field,  which was  found  to be
variable, with  a strength varying between -75~G  and +166~G, possibly
related to the spectroscopic variations.  Its spectral energy distribution (SED) presents
strong  near and mid-infrared  (MIR) excesses \citep{hillenbrand92},
as   well   as   a    weak   emission   at   millimetric   wavelengths
\citep{henning94}. This indicates that a circumstellar disk is present, which is  probably  of  low  mass  \citep[0.006~M$_{\odot}$;][]{siebenmorgen00}.
\citet{grady05} found no indication of a nebulosity at radii larger than 70~AU,
which was later confirmed by MIR spatially resolved observations
\citep{preibisch06}. The MIDI/VLTI 
measurements indicated  an angular extent of the  MIR emission smaller
than that of  other HAeBes, and were consistent  with an inclined disk
(i$\sim$58\dg,  PA$\sim$115\dg),  truncated  at  an  outer  radius  of
2.5~AU. The authors speculated that the low mass and the small outer radius of
the disk are probably caused by mechanisms of dynamic clearing by a close
binary.  In addition,  \citet{rodrigues09} measured an intrinsic
linear  polarization,  due to  scattering  of  the  stellar light  off
circumstellar material along PA$\sim$137\dg, in agreement with the 
PA of the disk measured by \citet{preibisch06}.

Very little is known about the structure and morphology of the circumstellar disk in the
first AU, as no spatially resolved observations have so far been published on  the  NIR emission.   Here,  we present  the  first
observational study of the circumstellar disk around \hr~at the sub-AU
scale using the AMBER/VLTI. We gathered a large number of measurements
in the  $H$ and $K$ bands  from 2008 to  2010, and present here  the first
reconstructed images of \hr, as  well as a qualitative model to account
for the interferometric measurements.  \\
The article is organized as follow.  Sect.~\ref{sec:datared} describes the observations, the
data    processing    and    the    image    reconstruction    method.
Sect.~\ref{sec:image} presents the reconstructed images 
obtained in the $H$ and $K$~bands, and Sect.~\ref{sec:model} provides a
temptative disk model.   In Sect.~\ref{sec:discussion}, we discuss our
results and conclude.  

%
%
% \begin{figure*}[!t]
%  \centering
% \begin{tabular}{cc}
%%  \includegraphics[width=0.47\textwidth]{imageK_2}
%% \includegraphics[width=0.47\textwidth]{HR5999_dirac_K_MiRA_img__Final_zoom10_DoubleAxis.pdf}
%  \includegraphics[width=0.47\textwidth]{imageK_6.pdf}
%  &
%  \includegraphics[width=0.47\textwidth]{imageH_6.pdf}
% \end{tabular}
%   \caption{\label{fig:image}  Reconstructed  images  in  the  $K$  band
%  (left)  and in  the  $H$ band  (right).   The interferometric  beams
%  (defined as the maximum angular resolution) are
%  given in the lower right corners. The images are shown on a 14 mas x
%  14 mas scale.}  
% \end{figure*}
%

 \begin{figure*}[!t]
  \centering
   \includegraphics[width=0.5\textwidth]{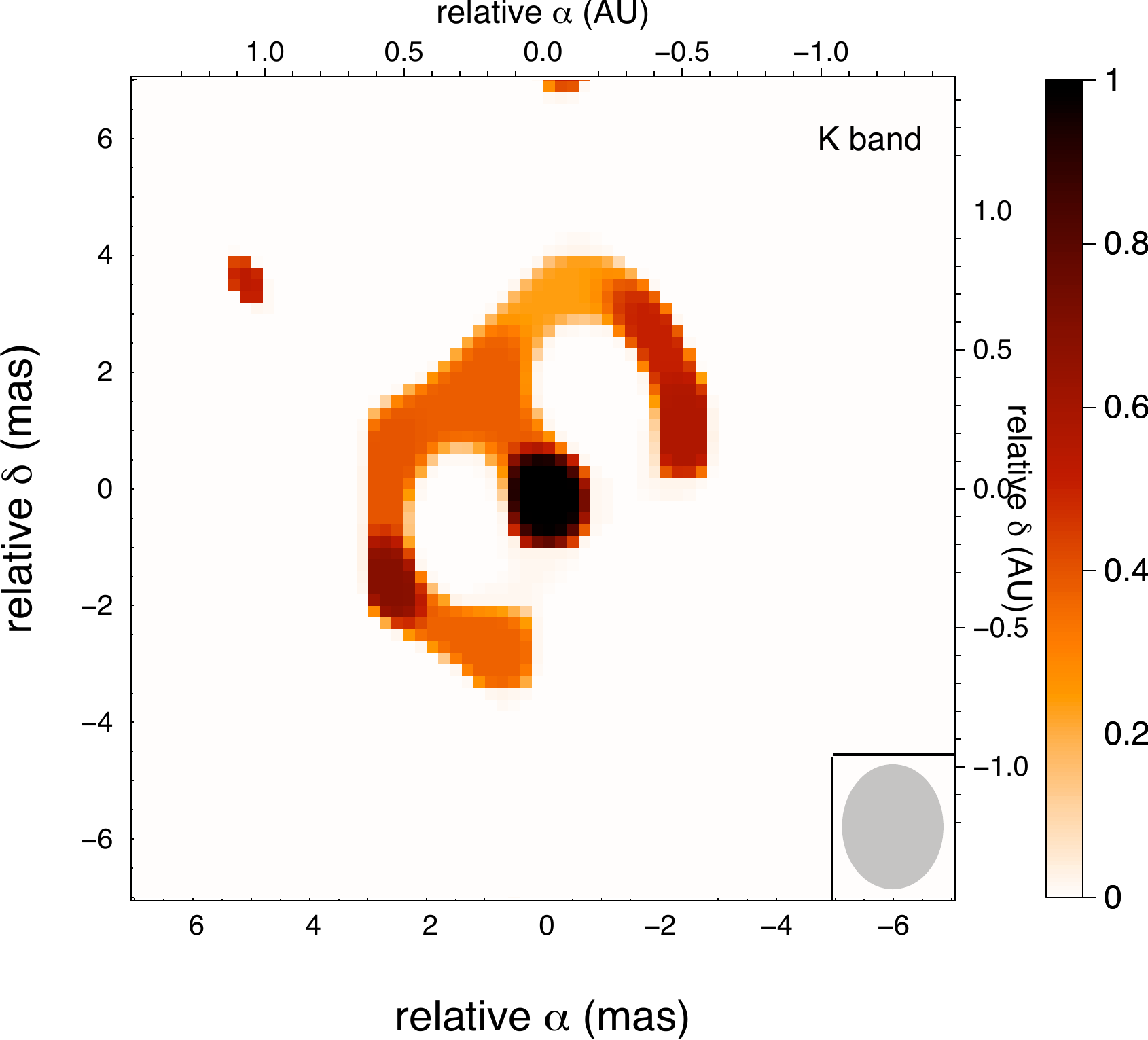}
    \caption{\label{fig:image}  Reconstructed  image  in  the  $K$  band.   The interferometric  beam
  (defined as the maximum angular resolution) is
  given in the lower right corner. The image is shown on a 14 mas x
  14 mas scale. The $H$  band image (not shown) brings little reliable information. } 
 \end{figure*}

\section{Observations and data processing}
\label{sec:datared}
\subsection{AMBER measurements}
We observed \hr~between February 2008 and June  2010, during fourteen
nights.  We  used the near-infrared  instrument AMBER, located  at the
Very Large Telescope Interferometer \citep[VLTI;][]{vlti1}. AMBER enables
the simultaneous combination of three beams  in the $H$ (1.69-1.73~\m) and $K$ bands
(2.0-2.4~\m),   with   a    spectral   resolution   up   to   $\sim$12~000
\citep{petrov07}. 
In  the following, we  present measurements obtained
with  the  low  spectral  resolution mode  ($\lambda$/$\Delta  \lambda
\sim$35). The data were obtained within programs of Guaranteed Time.
We performed these  observations using the relocatable 1.8~m
auxiliary telescopes (ATs) in seven different configurations, sampling
a large  range of baseline  position angles and providing  an excellent
\uv coverage.  The longest baseline is $\sim$128~m
corresponding to a maximum angular resolution $B/\lambda$ of 1.3~mas. A summary of
the   observations    presented   in   this   paper    is   given   in
Table~\ref{tab:obs}.  Each measurement for \hr~was encircled by observations of a calibrator
targets (HD136014, HD123004, HD145921, HD137730)  to measure the instrumental transfer function and
correct  for  instrumental effects.   About 20\%  of  the observations  were
performed using the fringe-tracker FINITO \citep{lebouquin08}.

The  data  reduction   was  performed  following  standard  procedures
described  in   \citet{tatulli07}  and  \citet{chelli09},   using  the
\texttt{amdlib} package, release 2.99, and the \texttt{yorick} interface provided by the
Jean-Marie                                                     Mariotti
Center\footnotemark{}\footnotetext{$\textrm{http://www.jmmc.fr}$}.
Raw spectral visibilities, differential phases, and closure phases were extracted for
all the  frames of each  observing file.  A  selection of 20\%  of the
highest quality frames  was  made  to  enhance the  signal-to-noise  ratio.   The
transfer   function   was  obtained   by   averaging  the   calibrator
measurements, after correcting for their instrinsic diameters.  
In general, the $H$ band data have a lower signal-to-noise ratio than $K$-band due to lower
flux  and instrumental  transmission. Consequently,  no  $H$-band data
could  be   retrieved  during  four   nights  (2009-04-08,  2010-06-07,
2010-04-10,  2010-03-21).  The  closest  detected companion  (Rossiter
3930 at 1.4'') is outside the field of view of the ATs ($\sim$300~mas), and does
not impact our measurements. \\
The total dataset covers observations over more than two years,
and consists of  1014 and 338 $K$-band visibilities  and closure phases,
respectively, as well  as 498 and 166 $H$-band  visibilities and closure
phases,  respectively.  The calibrated data in the OI-FITS format \citep{pauls05} are available upon request.
Figure~\ref{fig:uv}  gives the  resulting  \uv
coverages. The processed visibilities and closure phases are presented
in Figs.~\ref{fig:V2obs} and \ref{fig:CPobs}. A
large  scatter  in  the  data  can  be  seen,  in  particular  in  the
visibilities, that  is not  only due  to the fact  that we  sample the
object with  different baseline position angles (details  are given in
Sect.~\ref{sec:model}).

\subsection{Image reconstruction process}
According to the Van Cittert-Zernicke theorem, the complex \emph{visibility},
$V\,\mathe^{\mathi\,\phi}$, is the Fourier
transform of the object brightness distribution at the spatial frequency of
the  observations  ($\mathrm{B}/\lambda$; where  B is the  baseline  and
$\lambda$ the wavelength).  The visibility amplitude, $V$, is related to the
spatial extent of the emission, and the phase, $\phi$, to the location
of the photocenter. However, the absolute value of the phase is lost because of random atmospheric
perturbations.  By adding the phases of the fringes measured
for each  baseline over a  3-telescope configuration, one  can measure
the  \emph{closure   phase},  which  is   insensitive  to  atmospheric
disturbances \citep{jennison58, Monnier2003}. The closure phase
includes part of the Fourier phase information, and is related to the global
asymmetry of the emission: a point-symmetric object has a zero closure phase.
The main observables are therefore the squared visibility amplitudes, $V^{2}$,
and the closure phases (CP). The goal of the image reconstruction is to numerically retrieve an
approximation of the true  brightness distribution of the source given
a set of $V^{2}$ and CP. To account for the data, the Fourier transform
of the image should fit the measurements.  However,
because of the sparse \uv coverage, the image reconstruction problem is ill-posed
as there are more unknowns than measurements.  Additional constraints are therefore required to
supplement  the  available  data  and  retrieve a  unique  and  stable
solution.  %% Few algorithms exist  to perform image reconstruction, such
%% as  the  ones  presented  in  \cite{Hofmann1993,  Ireland2006,  baron,
%%   meimon}. 
In this paper, we use the Multi-aperture image 
Reconstruction   Algorithm  \citep[\Mira;   by][]{Thiebaut2008}.   The
solution  is normalized  and  positive  and the  code  finds the  closest
agreement between the  image and the data, favoring  e.g., the simplest or
the  smoothest image  \citep{Thiebaut2009}.  For this work, we tested three different regularizations (so-called 'total variation', 'MEM' (for maximum entropy method), and 'compactness') that led to very similar results. In the following, we show images obtained with the 'total variation' regularization that minimizes the norm of the image gradient and favors areas with steep but localized changes \citep{Strong03}. For more details of the
image reconstruction process, we 
refer the reader to \citet{renard10a} and \citet{renard11}. We used a
150x150 grid, with a scale of  0.2 mas/pixel, and the $V^{2}$ and CP
have  a  weight  inversely  proportional to  their  individual  error.
All spectral channels were combined to maximize the \uv coverage and we obtained two polychromatic images,
for the $H$-band and the $K$-band, assuming a grey emission within each filter.  \\

\section{Reconstructed images}
\label{sec:image}
We  present  in Fig.~\ref{fig:image}  the  resulting image in the $K$ band.   The interferometric  beam is shown  in the  lower right
corners, and is defined as the maximum angular resolution along the U
and V axis (1.8~mas x 2.3~mas in the $K$ band and 1.6~mas x 1.8~mas in
the $H$~band). Using MiRA (or a similar fitting procedure), one can
retrieve spatial information at about half the maximum angular
resolution   of   the   interferometer   (performing   the   so-called
'super-resolution').   The reconstructed  image  resolves  the
inner environment around HR~5999, and reveals several features.  The 'blobby' aspect of the
image is probably due to the incomplete \uv coverage. \\
The $K$-band image gives indications of a bright spot at the
center and an elongated structure in a ring-like shape.  The central spot is $\sim$1.8~mas
wide,  i.e., about  our  limit  in resolution.   Using  an ellipse  to
describe the ring-like structure, we find a major axis at the inner
edge   of   $\sim$5.5-6~mas   (i.e,   at   a   distance   of   210~pc,
$\sim$1.15-1.26~AU), a ratio [width/inner radius] of $\sim$25\%, an inclination of $\sim$40-50\dg, and an orientation of
the long axis along PA$\sim$130-140\dg. It provides 
$\sim$33-38\% of the total flux in the image (depending on the regularization), while the large central spot contributes to
the rest.  Between these two main  features, the image  reveals low or
zero emission. We interpret the elongated structure  as the disk
emission. The central spot is interpreted as the image of the star (of diameter $\sim$0.2~mas), and possibly of additional
unresolved  or partially resolved  circumstellar material  ($\sim$1 to
2~mas  wide). \\
\noindent  The $H$-band emission is more compact than the $K$-band emission, and the image is of quite low quality because the observations obtained are of much lower signal-to-noise ratio. The image consists of two blobs (separated by 3.8~mas, which we interpret as tracing the elongated structure  seen in the $K$-band image, i.e., a disk) and a central spot ($\sim$1.8~mas wide). These features have sizes about or slightly smaller than the maximum angular resolution, and while the separation between the two blobs is resolved at this resolution, the global $H$-band emission is only marginally resolved by  our observations. In  this case, the  reconstructed image provides only partial information \citep{lachaume03}, such as the orientation of the bulk of the emission, but cannot reproduce the structure of the circumstellar material emitting in the $H$ band. The large error bars in the $H$-band data lead to a much smaller dynamical range, probably preventing the ring to appear in the image. Finally, the chromaticity of the emission in the $H$-band has a strong impact on the  visibility (with a clear increase with B/$\lambda$, across the spectral channels of an observation; Fig.~\ref{fig:V2obs2}). Considering these caveats, the $H$-band image provides little reliable physical information, and will not be discussed further.
%
%, while their
%outer edges are $\sim$5.6~mas apart (corresponding to the inner radius of the
%K-band ring-like feature).  They are aligned along
%PA$\sim$130-140\dg.  These blobs contribute $\sim$20\% of the total flux
%in the image, the central spot provides the remaining. Inside the blobs, the image gives a low or null
%emission, a  shared characteristic with the $K$-band  image. 

%
%This requires a high contrast between the disk emission, and the region located inside: \textit{these images
%bring model-independent  evidence for a strong change  in the surface
%brightness within the first AU surrounding \hr. }

\vskip 0.1cm
Considering the incomplete  \uv coverage, the limited angular resolution, the large uncertainties, and the scatter in the observations, great caution is required when interpreting the $K$-band image.  Interpreting the exhibited details but also the hidden features in these images is not straightforward, as the  reconstruction process produces artefacts.  We performed numerous tests to understand this issue. Our tests show that the images depend on the availability and precision  of measurements at  high  spatial frequency  (i.e., at long  baselines), and whether these measurements  have a high enough angular resolution to discriminate, e.g., a ring from a uniform disk.  \\
\indent  The ring in the $K$-band image ($\sim$6 mas wide) is resolved at the angular resolution of our measurements, and its appearance is systematic in all the tests we performed with different regularizations. On the other hand, the central spot has a size similar to the maximum angular resolution achieved with our observations, and is only marginally resolved. The exact morphology of this emission can therefore not be determined by the image. 
The combination of the different features leads to a complex visibility curve that depends on their morphology but also on their relative flux contributions. In addition, possible incorrect calibration of the absolute visibility as well as the chromaticity of the emission (that we consider grey within each band for the image reconstruction) can produce an additional scatter (up to 5-10\%), adding complexity to the visibility curve. However, the ring is encoded in the differential visibilities (corresponding to different spatial frequencies)  that are not affected by any calibration issue as well as in the closure phases and in the observations at very long baselines that are well-accounted for by MiRA. Given the large amount of data, the ring is systematically retrieved in the reconstructed images.  Figs.~\ref{fig:V2mira},~\ref{fig:CPmira} show the best-fit to the $K$-band observations obtained with MiRA ($\chi_{r,V^{2}}^{2}=11.2$; $\chi_{r,CP}^{2}=0.6$), and corresponding to the reconstructed image in Fig.~\ref{fig:image}. One can see that some of the scatter in the observations is well reproduced indicating a complex structure resulting from a combination of resolved features (such as the ring) and marginally-resolved features (such as the central spot).  \\
\indent We must also consider the sensitivity of the image reconstruction algorithm to  a low brightness surface such as the one inside the ring-like feature, a common feature of theses images.  The scatter in the   measurements at  a given  spatial frequency has the 
consequence of lowering the signal-to-noise ratio (since at a given
spatial frequency, different values of the visibilities and CP have to
be reproduced for a single image value) and producing 
a lower dynamical range.  In practice, this decreases our capability of
detecting low surface brightness emission, and favors sharp transitions in the
image.  On the other hand, if \hr~were surrounded by a uniform disk, the reconstructed image would have shown an elongated disk-like structure and not a ring-like feature.  It is therefore likely that \hr~disk displays a strong change in the surface brightness within the first AU. 

To more clearly understand  the image and avoid over-interpretation,  we attempt  to apply a fitting procedure  to the  V$^{2}$ and  CP based  on disk  models that we  develop in  the next
section.

%In particular, in the $H$-band, the data at the longest baseline do not reach this regime (see,
%Fig.~\ref{fig:V2obs}, right), with visibilities  still at a high level
%(V$^{2}\sim$0.2).  This indicates that the $H$-band  emitting region is
%more compact than the one emitting in the $K$-band, and that it is only partially
%resolved by  our observations. In  this case, the  reconstructed image
%only provides \textit{partial} information \citep{lachaume03}, such as the orientation and elongation of the bulk of the emission, but can not reproduce the
%exact structure of the circumstellar material emitting in the $H$ band.
%In the K-band, the data at the longest baseline is close to this regime and reveal reliable information such as the size of the emission and its orientation. 

\section{Disk model}
\label{sec:model}
The main features seen in the images  of \hr, i.e., a star, a disk, and
a lower surface brightness inner region resemble the observed
properties of  the circumstellar  matter around other  Herbig~Ae stars
\citep[e.g.,][]{isella08, eisner09, benisty10}.  To understand the images, we
propose a  disk model that qualitatively  reproduces the observations. \\
As shown in Figs.~\ref{fig:V2obs},~\ref{fig:V2obs2}, the squared
visibilities smoothly decrease with the baseline length, and reach
values of $\sim$7\%  (in $K$) and $\sim$17\% (in  $H$) for the longest
baseline. The closure phases (Fig.~\ref{fig:CPobs}) are close to zero~when measured with short-baseline configurations, and increase up
to $\sim$20\dg~for long-baseline configurations.  Both 
visibilities  and closure  phases  change with  the baseline  position
angle and hour angle of the observations.

\subsection{Spectral energy distribution}
Photometric   measurements   were    gathered   in   the   literature
\citep[2MASS;][]{hillenbrand92, dewinter01}, enabling us to construct a SED
(Fig.~\ref{fig:sedv2},  left).    Using  a  Kurucz  model   for  a  A5
photosphere at T$_{\rm{eff}}$=8000~K and Av=0.49, we find a stellar
contribution to the $K$ and $H$ fluxes of $\sim$22\% and $\sim$40\%, 
respectively.  The adaptive  optics study by \citet{stecklum95} showed
that the flux of the companion, Rossiter 3930, is much smaller than \hr~up to NIR wavelengths (i.e., $\Delta H, K$ = 3.1, 3.6 mag), and
can be neglected in the following.

\subsection{Models}
\textbf{A uniform ring?}
We first fitted geometric models to our visibilities to estimate the main
characteristics  of  the  NIR  emission  and check  whether  they  are
consistent with the image features.  We used a uniform ring model,
with  a ratio of width to inner  radius  of  $\sim$20\% to  account for  the
emission of the  inner parts of a circumstellar  disk.  In the $K$-band,
we find  that the  best fits  are achieved with  an inner  diameter of
$\theta=$4.4$\pm{0.2}$~mas (i.e.,  an  inner radius  of $\sim$0.45~AU),  an
inclination of i=48$\pm{11}$\dg, and an orientation of PA=116$\pm{11}$\dg. In
the $H$-band,  the best-fit models give  $\theta$=3.6$^{+2.4}_{-0.9}$~mas, i=42$^{+26}_{-42}$\dg,
and PA=121$^{+59}_{-121}$\dg. These models provide an unsatisfactory fit to the
V$^{2}$,  with large values of reduced $\chi^{2}$, which probably reflects the more
complex  circumstellar  environment  and is possibly caused by  the  large
scatter in the visibilities.  Their predictions however give values of
i     and    PA that     agree   with     MIR    \citep[i=58\dg,
PA=115\dg;][]{preibisch06}      and      polarimetric     measurements
\citep[PA=137.2\dg;][]{rodrigues09}, and with the NIR reconstructed images.\\

\begin{figure*}[t]
 \centering
\begin{tabular}{ccc}
 \includegraphics[width=0.32\textwidth]{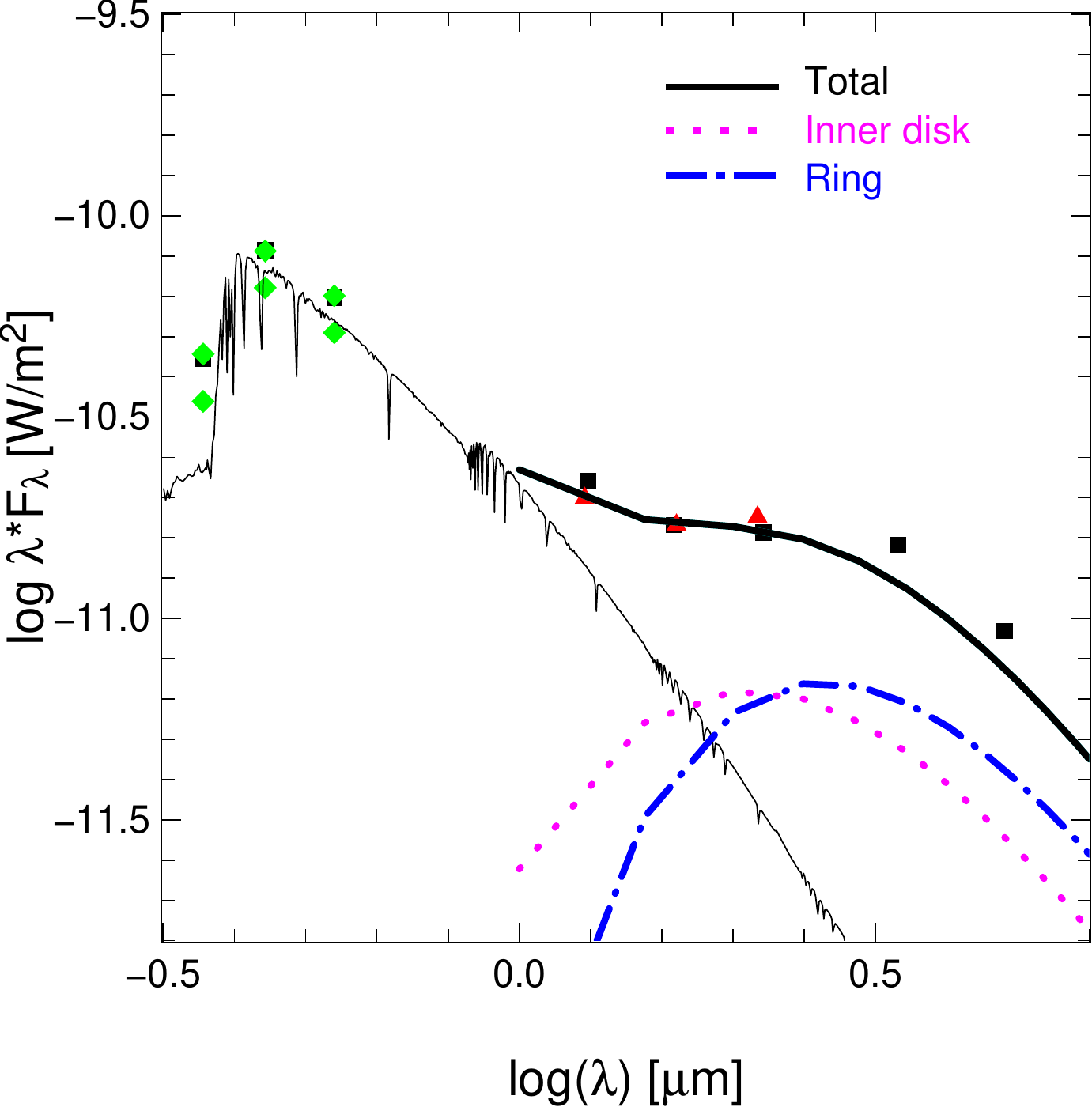}
 &
 \includegraphics[width=0.32\textwidth]{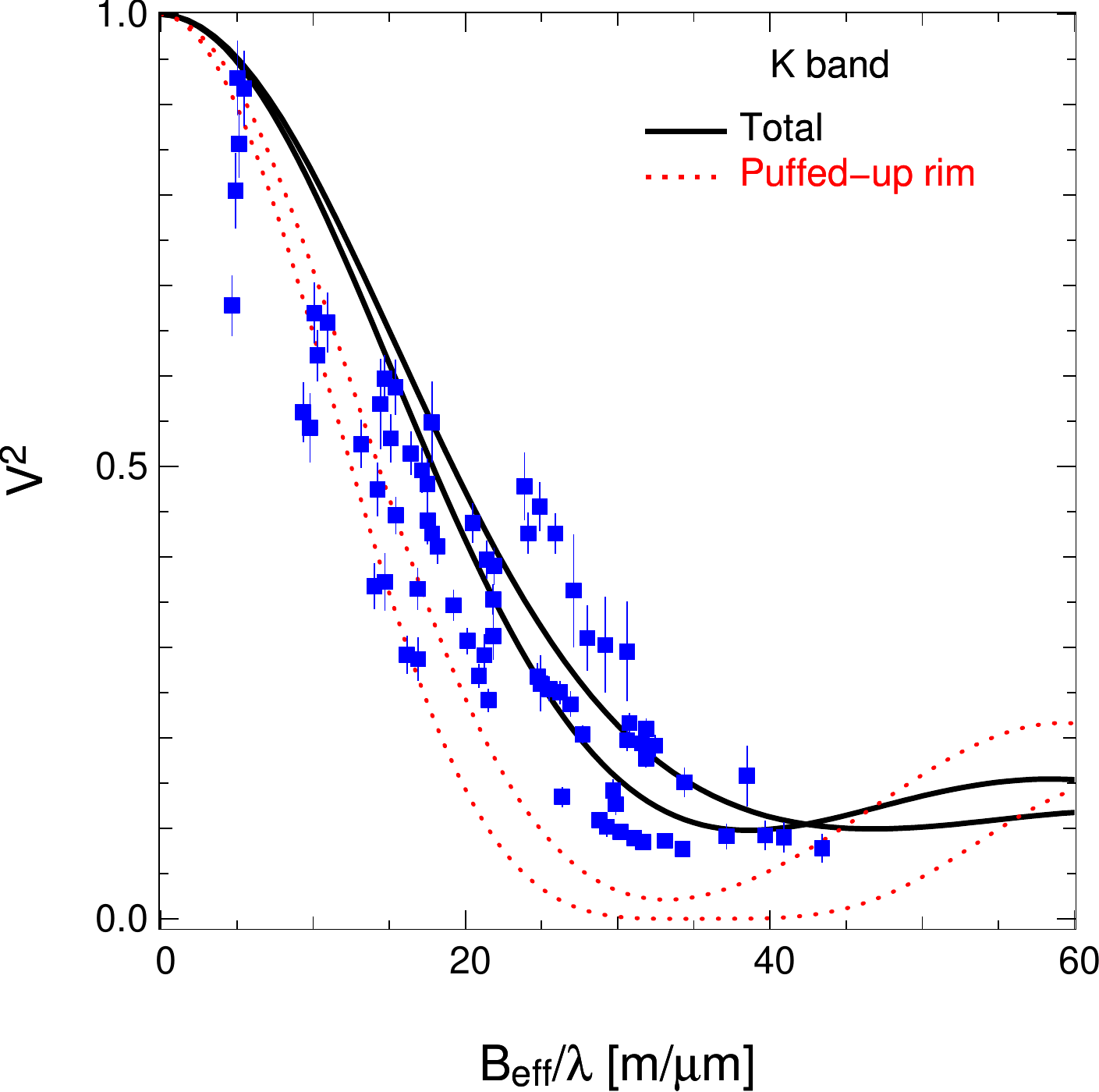}
& 
 \includegraphics[width=0.32\textwidth]{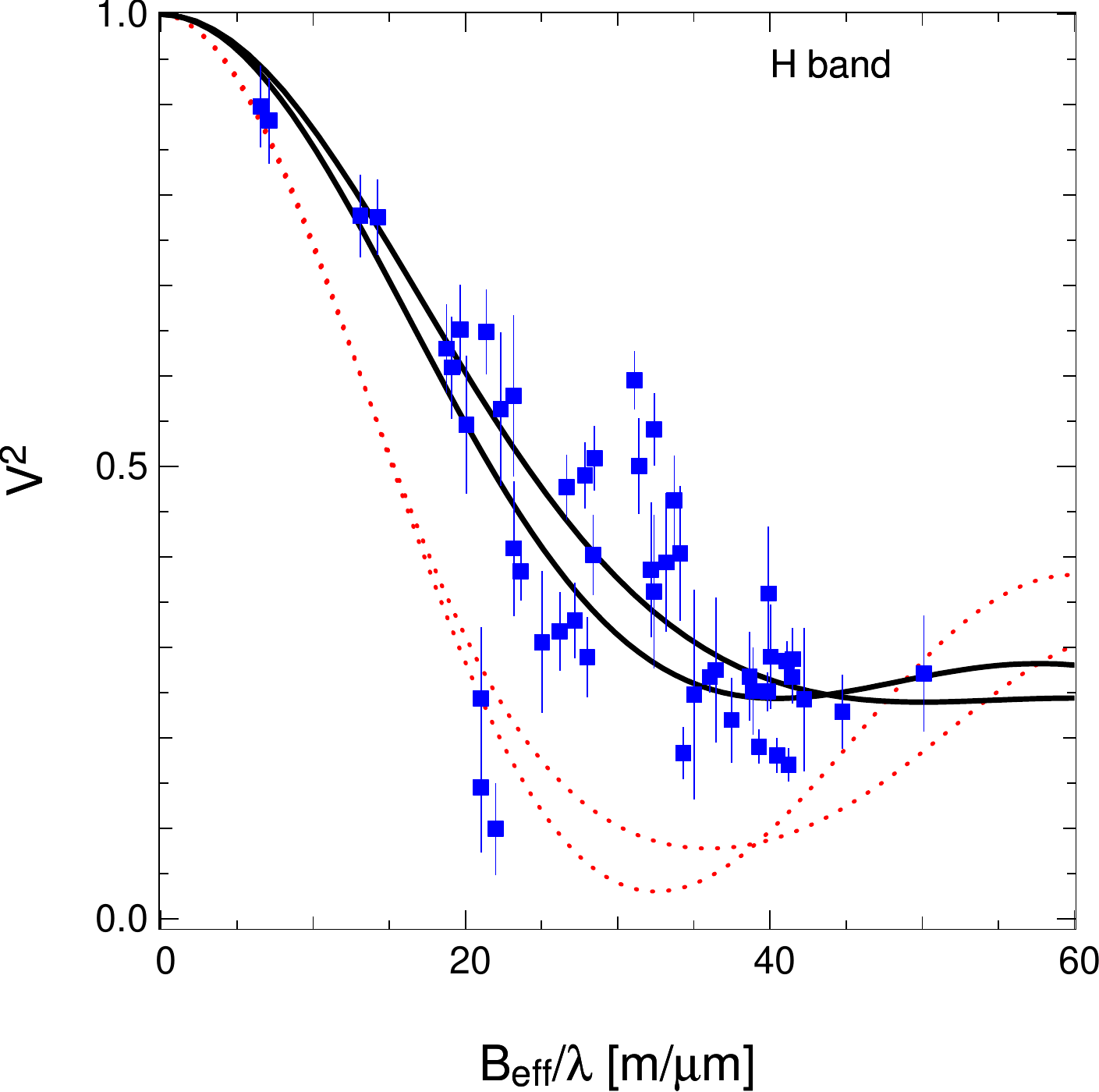}
\end{tabular}
  \caption{\label{fig:sedv2} The black  curves give the predictions of
 the two-component disk model  for the SED (left), the broad-band
 visibilities   in  the   $K$-band  (middle)   and  in   the  $H$-band
 (right). The  visibilities predicted by  the puffed-up rim  model are
 added (red dashed curves). In both cases, the visibilities curves are shown along two perpendicular baseline position angles, against effective spatial frequency (computed with i=48$^\circ$ and PA=135$^\circ$). See the text for more details on the effective baselines B$_{\rm{eff}}$, and note that their values are smaller than the \textit{physical} baseline lengths given in Figs.~\ref{fig:V2obs} and \ref{fig:V2mira}. }
\end{figure*}

\noindent \textbf{A puffed up rim?} 
Early NIR interferometric  studies of  Herbig~Ae stars  have shown
that standard  accretion disks, extending  up to the  dust sublimation
radius do not  fit the observations and that  better fits are obtained
assuming  that the  disk  develops  a curved  rim  controlled by  dust
sublimation     and     its      dependence     on     gas     density
\citep{isella05,tannirkulam07}. The  detection  of non-zero  closure
phases supports the idea  that an asymmetric brightness distribution -
possibly a disk inner rim - contributes to the NIR emission.  We
therefore started by  examining a 'star + puffed-up  rim' to model our
observations, following \cite{isella05}.  We adopt the stellar properties d=210~pc, L=87~L$_{\odot}$,
T$_{\rm{eff}}$=7925~K and A$_{\rm{v}}$=0.49~mag.  The disk is  assumed to
be  in hydrostatic equilibrium.  The dust  consists of  silicates with
optical  properties given  by  \cite{weingartner01}.  The  evaporation
temperature, around  1500~K, depends  on the local  gas density  as in
\cite{pollack94}.  Since  the shape  of the rim  is controlled  by the
largest grains, we consider a single size for the silicate dust, which
therefore is the only free  parameter in the model.  The dependence of
the evaporation temperature on $z$  implies that the distance from the
star at which dust evaporates  increases with the altitude, i.e., that
the rim is curved.  \\ 
We find that our measurements  are inconsistent  with  a rim  only,
regardless of its location.  Using 1-micron large silicate grains with
a high cooling efficiency, we
find  that   they  sublimate   at  1486~K  at   a radius  of
$\sim$0.65~AU. With an inclination of 48\dg~and a PA
of 135\dg,  the rim accounts  for $\sim$62\% and $\sim$41\%  of the  $K$-band and
$H$-band  fluxes,  respectively, and  a  sharp  visibility profile  with
baseline that is  inconsistent     with     our     observations. We show in Fig.~\ref{fig:sedv2}, middle and right panels, the broad-band visibilities in the K and H bands respectively, against the effective spatial frequency B$_{\rm{eff}}$/$\lambda$. The broad-band visibilities were obtained by averaging the visibilities over the available spectral range.  
The effective  baseline B$_{\rm{eff}}$  is defined following  \citet{tannirkulam} as
$$B_{\rm{eff}}=                B\sqrt{cos^{2}(\theta)                +
 cos^{2}(i)sin^{2}(\theta)},$$
where $\theta$ is the angle between the baseline direction and the major
axis of  the disk, and $i$  is the disk  inclination.  
This representation is useful to show a large dataset in a concise way once the
inclination and position angle of the disk are known, as effective baselines account for the decrease in interferometric resolution due to the inclination of the disk in the sky. We note that $ B_{\rm{eff}} \leq B$.
Following \citet{tannirkulam}, a model-independent estimate of the disk inclination and position angle can be retrieved  by minimizing the variance of the observed visibility curve at a given baseline length. Using this method, we find that i=52$\pm$11\dg and PA=143$\pm$11\dg. Considering the large error bars, in the following, we assume that i=48\dg and PA=135\dg.
As it can be seen in the red dashed curves, the rim model with an inclination of 48\dg~and a PA
of 135\dg~cannot  reproduce the $H$
and $K$-band visibility measurements, and results in closure phases that
are  much higher than  observed (up  to $\sim$60\dg).   This indicates
that the actual level of asymmetry is lower 
than the one predicted by a puffed-up rim at 48\dg~of inclination, and that the bulks of the $H$
and of the $K$-band  emissions are  unlikely to originate in the same
physical region. \\
\indent The model of a rim of dust grains directly irradiated by the star
cannot fit our data.  However, we find that, for various dust
grain properties, the silicate sublimation occurs around 
$\sim$0.65~AU, a distance similar to the sharp
transition in surface brightness seen in the
$K$-band image.  We therefore interpret the ring-like feature in the $K$-band image as
tracing  the  emission  from   dust  at  $\sim$1500~K,  i.e.,  at  the
transition radius where silicate~condensate.\\

\begin{figure*}[t]
 \centering
\begin{tabular}{cc}
 \includegraphics[width=0.45\textwidth]{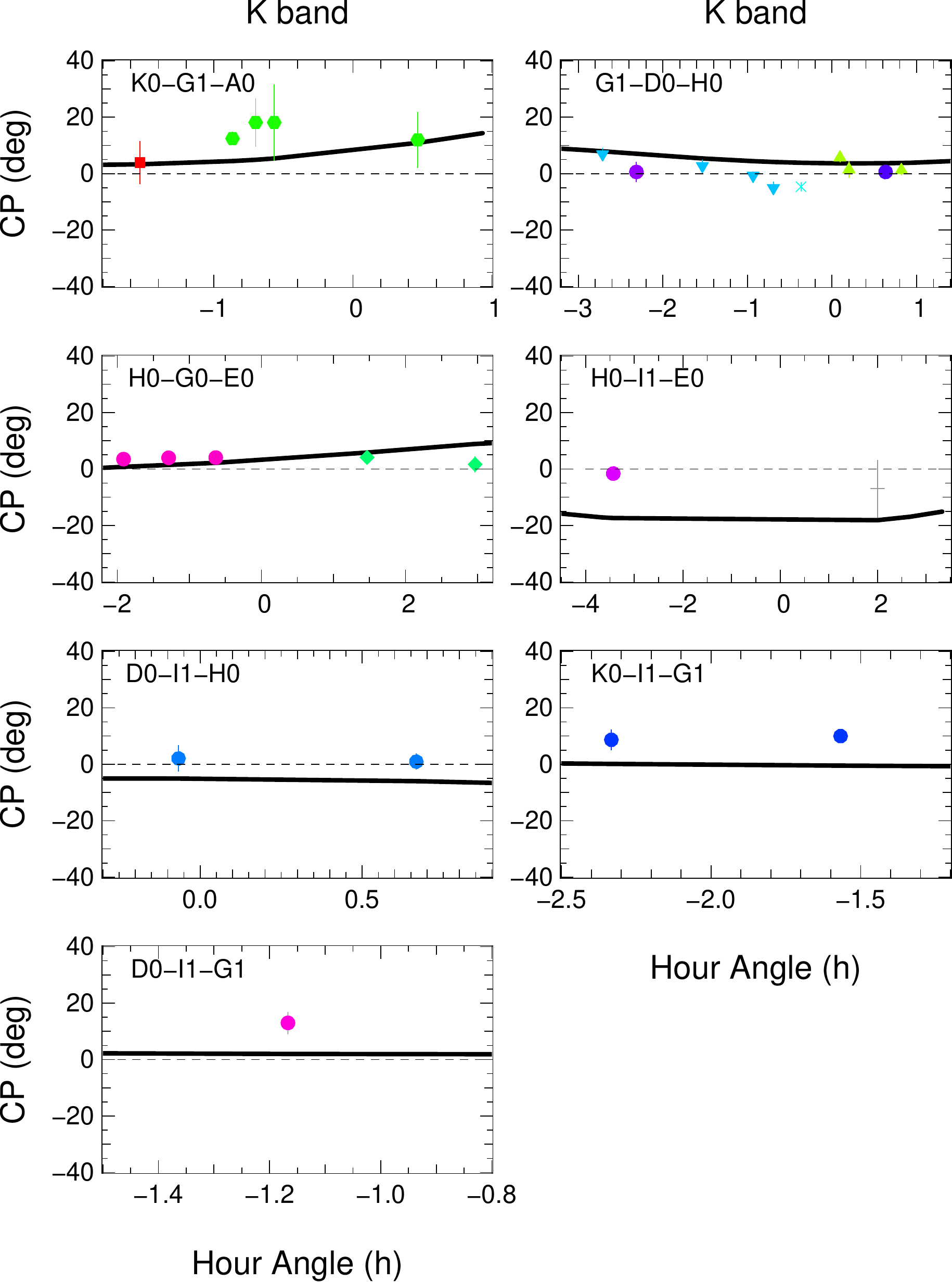}
 &
 \includegraphics[width=0.45\textwidth]{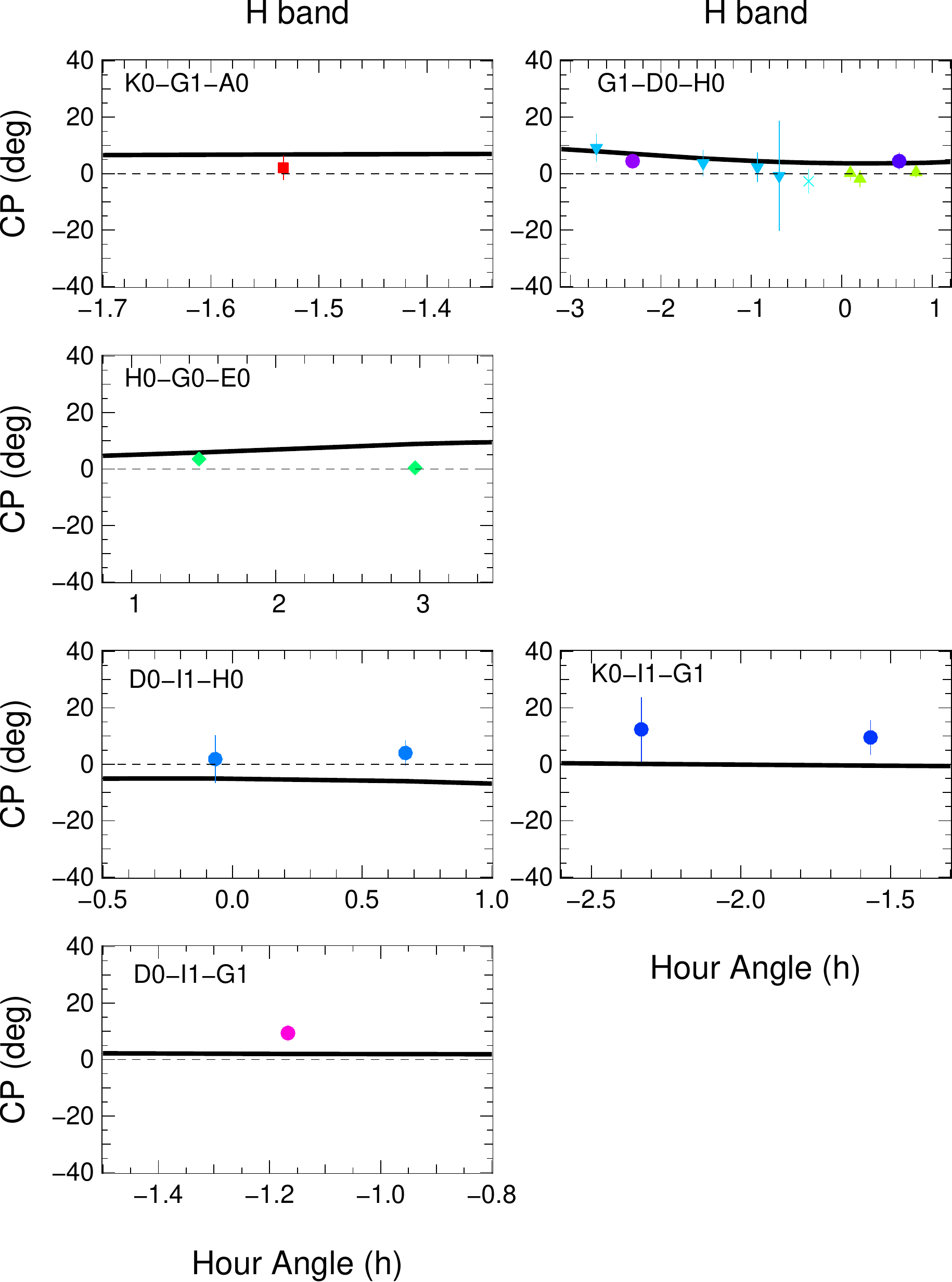}
\end{tabular}
  \caption{\label{fig:2}  The black  curves give the predictions of
 the two-component disk model for the broad-band closure phases in the
 $K$-band (left columns) and in  the $H$-band (right columns). Each panel
 corresponds  to a  different telescope  configuration, and  the color
 code is the one of Figs.~\ref{fig:V2obs} and \ref{fig:CPobs}.} 
\end{figure*}

%
% \begin{figure*}[!t]
%  \centering
% \begin{tabular}{cc}
%%  \includegraphics[width=0.47\textwidth]{HR5999_dirac_MiRA_img2colorsZOOM__Final.pdf}
%%\includegraphics[width=0.47\textwidth]{HR5999_dirac_MiRA_H+K_img2colors__Final_zoom5_DoubleAxis.pdf}
%\includegraphics[width=0.47\textwidth]{HR5999_dirac_MiRA_H+K_img2colors__Final_zoom7_DoubleAxis.pdf}
%  &
%%% \includegraphics[width=0.47\textwidth]{HR5999_ModFinal_dirac_MiRA_img2colorsRotZOOM_DispFact_3_test1.pdf}
% % \includegraphics[width=0.53\textwidth]{newImg.pdf}
%%\includegraphics[width=0.47\textwidth]{Img_22Oct_test4.pdf}
%%\includegraphics[width=0.47\textwidth]{imageHK_27Oct_test6H_test5K_zoom.pdf}
%\includegraphics[width=0.47\textwidth]{imageHK_27Oct_test6H_test5K_zoom7_2.pdf}
% \end{tabular}
%   \caption{\label{fig:imagemodel}  Reconstructed   images  in  two
%  colors ($K$ band in red; $H$ band in green), from the measurements (left)
%  and from  the model  (right), with the  original model in  the upper
%  left corner. The images are shown on a 14 mas x 14 mas scale.} 
% \end{figure*}

 \begin{figure*}[!t]
  \centering
 \begin{tabular}{ccc}
\includegraphics[width=0.33\textwidth]{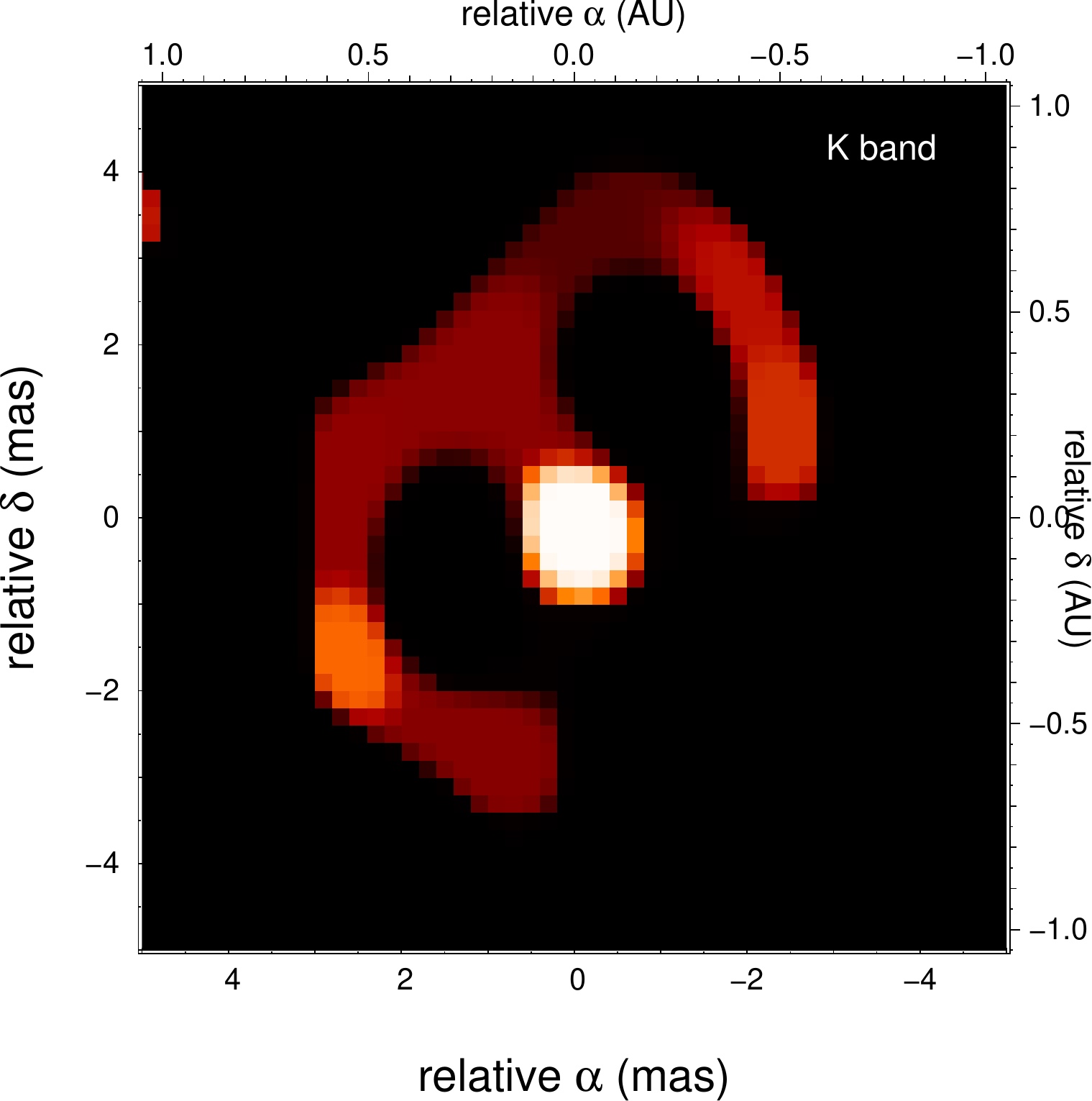}
%HR5999_dirac_MiRA_H+K_img2colors__Final_zoom7_DoubleAxis.pdf}
  &
\includegraphics[width=0.33\textwidth]{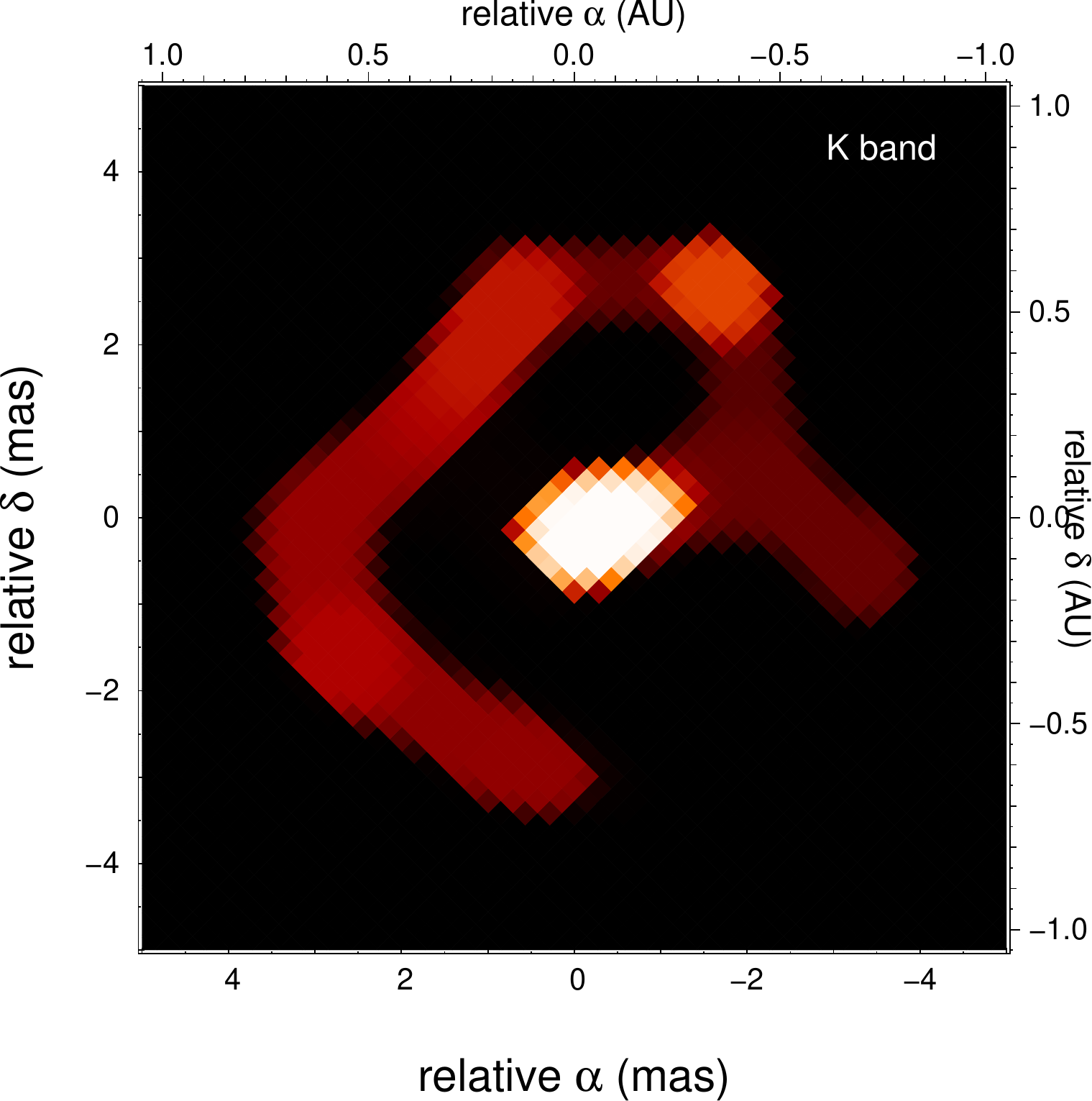}
%imageHK_27Oct_test6H_test5K_zoom7_2.pdf}
   &
%\includegraphics[width=0.33\textwidth]{HR5999_Fig5_c_heat}
%Model_K.pdf}
\includegraphics[width=0.33\textwidth]{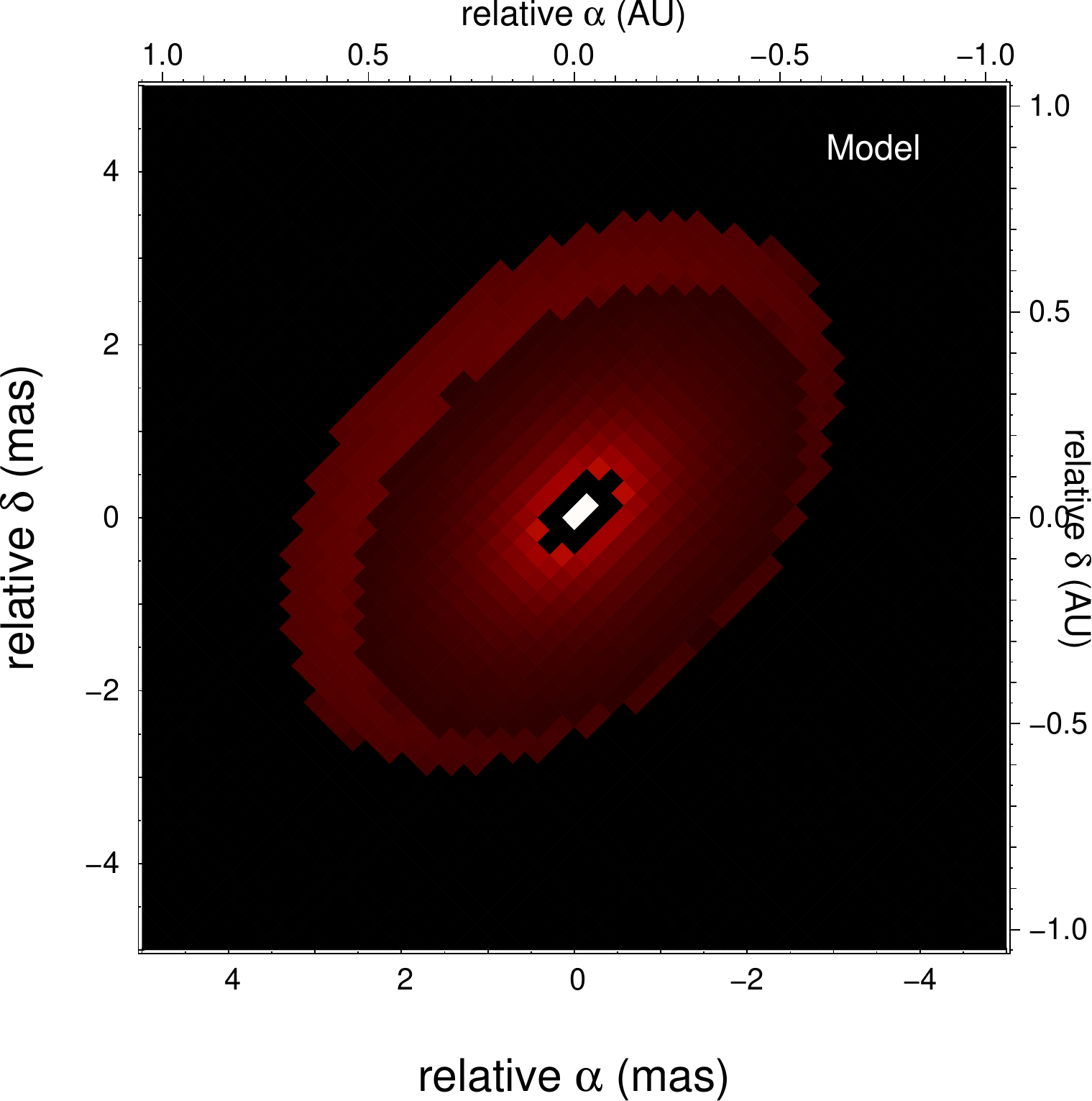}

 \end{tabular}
   \caption{\label{fig:imagemodel}  The $K$-band  image is shown, as reconstructed from the AMBER measurements (left), and as reconstructed from the simulated data from the model (middle). The latter is shown in the right panel. The images are shown on a 10 mas x 10 mas scale.}
 \end{figure*}

%
% \begin{figure*}[!t]
%  \centering
% \begin{tabular}{ccc}
%\includegraphics[width=0.33\textwidth]{HR5999_Fig5_a.pdf}
%%HR5999_dirac_MiRA_H+K_img2colors__Final_zoom7_DoubleAxis.pdf}
%  &
%\includegraphics[width=0.33\textwidth]{HR5999_Fig5_b.pdf}
%%imageHK_27Oct_test6H_test5K_zoom7_2.pdf}
%   &
%%\includegraphics[width=0.33\textwidth]{HR5999_Fig5_c_heat}
%%Model_K.pdf}
%\includegraphics[width=0.33\textwidth]{HR5999_ModFinal-papier2_imgModele}
%
% \end{tabular}
%   \caption{\label{fig:imagemodel}  Left: the reconstructed  image is represented in  two
%  colors  for the $K$ band (red) and the  $H$ band (green). Middle: same for the image reconstructed from the model that is shown in the right panel. The images are shown on a 10 mas x 10 mas scale.}
% \end{figure*}

\noindent\textbf{A two-component disk?} 
The previous simple models suggest that the $H$-band emission is more compact
than  the  $K$-band emission,  and  indicate  that  there is  material
located  inside the  silicate sublimation  radius, as already
found in other Herbig~Ae stars.  Assuming that the ring-like
feature in the  $K$-band image is related to  silicate sublimation, we
compose a model made of three elements: a star, a ring at the silicate
sublimation radius, and an inner disk. This 
inner and compact emission is expected to modify the shape of the visibility
curves and to smooth the closure phases predicted by a very asymmetric
component such as the puffed-up rim. In
this section, we  attempt to derive the main  characteristics of these
three  components, i.e., their  extents and  contributions to  the NIR
emission.  

Since it is unclear whether a rim would puff up  in the way computed in
\citet{isella05}  if inside  matter is  blocking part  of  the stellar
emission,  we  refer   to  the  inner  edge  of   the  dusty  disk  as
\textit{ring} instead of \textit{rim}. Determining its exact structure
is beyond the scope of this paper.  This ring traces dust condensation and
provides some asymmetric emission, as indicated by the non-zero closure
phases.  We compute its emission using the rim model at
R$_{\rm{sub}}$=0.65~AU, but where its luminosity is treated as a free parameter to 
enable SED fitting with an additional inner component.  We describe the inner disk using a radial temperature profile
T$\propto$r$^{-\alpha}$ as expected in a circumstellar disk, and a vertical optical depth~$\tau$.  \\
\indent We find an acceptable fit to the SED, the visibilities, and the closure
phases in the $H$ and $K$ bands using a model where the inner disk extends
from      R$_{\rm{in}}$=0.1~AU       to      R$_{\rm{sub}}$,      with
T$_{\rm{in}}$=2400~K, $\alpha$=0.4, and a vertical optical depth $\tau\sim0.4$. In this 
model, the ring contributes 40\% of the $K$-band flux, while the inner
disk provides  38\% of it.  In  the $H$-band, the  ring is responsible
for 26\% of the emission, the  inner disk for 34\%, leaving the star as
the major contributor  at 1.6~$\mu$m. The parameters of  the model are
summarized in Tab.~\ref{tab:bestmodels}.  %We use a radial temperature profile T $\propto$ r$^{-\alpha}$, with $\alpha$=0.4,
%T$_{\rm{in}}$=2400~K  as  the  temperature  at R$_{\rm{in}}$,  considering  a
%vertical optical depth $\tau\sim0.4$.  
We show  in Figs.~\ref{fig:sedv2} and~\ref{fig:2} the model predictions
 for  the SED,  the broad-band visibilities  and closure
phases  (full black lines).  By  spreading the  NIR emission across  a broader range  of radii
(compared to the puffed-up rim), i.e., from R$_{\rm{sub}}$ to R$_{\rm{in}}$, the shape of the
visibility-versus-baseline curve  is well reproduced and  the high closure
phases predicted by  the model of the puffed-up  rim are smoothed out,
resulting in values close to the observations (from 0 to 20\dg). 
A temperature gradient within the inner disk is needed to reproduce \textit{both} the $H$ and $K$ band visibilities together, as a single temperature disk at a specific $\tau$ cannot.  The model is shown in Fig.~\ref{fig:imagemodel}, right. 
Considering the large scatter in the observations, we do not claim the
uniqueness of the parameters of our 
model ($\tau$,  T$_{\rm{in}}$,  $\alpha$),  although  they  provide  a
qualitatively good fit to the observations, and the extents and flux
ratio of each component are well constrained and in agreement with
the images. 

\begin{table}[b]
\centering
\caption{\label{tab:bestmodels}
  Parameters of our model with inner disk  between R$_{\rm{in}}$ and
  R$_{\rm{sub}}$, where silicates sublimate.  For both bands, the
  ratio  of the
  flux   contributions  from the star (F$_{*}$), the ring at dust sublimation(F$_{\rm{sub}})$,   and   the inner disk (F)   to   the    total   flux   in   the   \textit{model}  (F$_{\rm{tot}}$) are reported. The inner disk has a vertical optical depth $\tau \sim 0.4$, and the exponent of the radial temperature profile is $\alpha$=0.4. } 
\begin{tabular}{c|cccccccc}
\hline
\hline
Wavelength   &    R$_{\rm{sub}}$&   R$_{\rm{in}}$   &   T$_{\rm{in}}$&
F$_{*}$/F$_{\rm{tot}}$&
F$_{\rm{sub}}$/F$_{\rm{tot}}$&F/F$_{\rm{tot}}$\\ 
 & [AU]& [AU] & [K]& [\%] & [\%]& [\%]\\ 
\hline
\hline
$K$ band & 0.65 & 0.10 &  2400 & 22 & 40 & 38\\ 
$H$ band & ''& ''&  ''&  40 & 26 & 34\\
\hline
\hline

\end{tabular}
\end{table}

\subsection{Model image}
Our model produces a strong variation in surface brightness in the first AU, 
from the star to the ring.  To better interpret the reconstructed image presented in
Fig.~\ref{fig:image},  we performed an  image reconstruction  from the
visibilities and closure phases of our model.  To do so, we
computed synthetic data sets from the model image, with an identical \uv
coverage as the  observations, the same errors, and  a similar scatter
in the visibility 
measurements.  We  present an example of a reconstructed  image from the
model  in  Fig.~\ref{fig:imagemodel}, middle,   compared  to  the  real  image (left).  The image reconstructed from the model shares the
same  characteristics  as  the   real  image,  i.e.,  an  incomplete
ring-like feature  in the  $K$-band oriented along  PA$\sim$135\dg~and
inclined by $\sim$45\dg, with an inner diameter of $\sim$5.5~mas, and  an  extended central  spot.   This  confirms  that our  model  provides a qualitatively good description of our data.  In particular, the missing part  of the ring-like
feature can  be explained  by the low-brightness  edge of  an inclined
rim (while the other edge is much brighter).  It also shows that although the 
star is unresolved 
in the model with a diameter of $\sim$0.2~mas (i.e., inside the central pixel), the
central spot in the reconstructed image is much more extended ($\sim$2.1~mas; similar to the
real image).  Its size is about the interferometric beam size, but in the
model image, it has much more flux than the star's, due to the inner disk 
emission on unresolved scales as small as 0.1~AU. This could indicate that the
central spot in the real image also includes an additional emission to
the star's. \\
\indent Several tests have led us to conclude that the scatter in the visibility measurements lowers the achievable dynamics which
can result  in an image that has  low or no emission  inside the ring.
In practice, in the $K$-band model, 
the ring is responsible for 40\% of the 
total flux, i.e., each  pixel on the ring provides (on average)
$\sim$0.10\% of it (corresponding to 0.25\% per squared mas) and can be retrieved. The inner disk contributes 38\%
of the flux, which means that each of the pixels
tracing  the inner  disk provides  (on  average) $\sim$0.06\%  of the  total
$K$-band emission (i.e., 0.15\% per squared mas), and is not necessarily
retrieved.  %{\bf Quantify the 
%dynamics/limit of MiRA in units of surface brightness?}. 
The  reconstructed image  from  the model  (Fig.~\ref{fig:imagemodel},
middle) shows that, in our case, the process is probably
able to retrieve the ring emission but not to reproduce the inner disk
emission. \textit{Precise} data at higher angular resolution would be needed to
do so. The reconstruction process
gathers most of the 
inner disk emission  together with the star in  the central spot, that
contributes in the real image to 62\% of the $K$-band emission.  \\
\indent The $H$-band image reconstructed from the model is similar to the
real one, with an elongated structure extending up to the ring radius,
and  confirms that  the $H$-band  emission  is more  compact and  only
partially resolved by our measurements.

%% Once the parameters of the image reconstruction process are set, the images
%% can be reconstructed. The solutions are not straightforward to analyze because of artifacts caused by the image reconstruction process or the
%% quality of the data set, e.g., voids in the \uv plane, error bars. There are
%% no objective criteria to distinguish between the actual structures from the
%% object and the artifacts caused only by the data structure. We therefore
%% performed a comparative analysis between our results and the results obtained
%% from simulated data from the B10 model for mwc. To do so, we simulated fake
%% data sets, using the B10 model and the same \uv plane and errors as
%% in the real data set. Image reconstruction was also performed for the simulated
%% data in the same conditions and compared to the image model. This comparative
%% method is important to understand what could be trusted in the actual
%% reconstructed images and what could not. 
%% Goodness of fit de l'image pour les CP et les V2 + 

%
%\textbf{Power of image reconstruction that can reveal features around a marginally-resolved source that would not have been seen by inspecting only visibility curves.}

\section{Summary \& conclusions}
\label{sec:discussion}
We have presented images  of the  close circumstellar environment  of HR~5999  reconstructed from  extensive interferometric datasets and the MiRA algorithm, assuming a gray emission within each band.  The K-band image shows two clear features: a marginally-resolved central spot and a resolved ring-like feature with an inner diameter of $\sim$6~mas. The H-band emission is more compact and the corresponding image suffers from observations obtained at low signal-to-noise ratio. With the maximum angular resolution reached by the VLTI baselines ($B/\lambda \sim$1.3~mas), we are unable to probe the exact structure of the inner disk, the corresponding H-band image providing limited and partial information, such as the orientation of the emission. 
To avoid mis-interpretation of the images, we have used a physical model of an unresolved star, that is surrounded by an inner disk of low surface brightness inside a ring located at the silicate sublimation radius.
In the model, the inner disk contributes 38\% and 34\% of the $K$ and  $H$-band flux, respectively, and the ring  does to the levels of 40\% and 26\%, respectively.  As they both contribute similarly to the NIR
emission but are distributed on different ranges of radii, one requires a
strong surface brightness contrast at $\sim$0.65~AU.  We interpret this as being caused by silicate condensation in a low optical depth region that implies a sudden
increase in opacity. The properties  that we derive for HR~5999 are very similar to the cases of e.g., AB~Aur, HD~163296, and MWC~758.  If we speculate that the low surface
  brightness traces  a low density region in  an optically thick
  dusty disk, it is interesting to speculate whether this is a general feature of Herbig~Ae stars.  The NIR emission was
originally  thought to arise  from the  inner edge  of the  dusty disk
only, and that  a similar (or  a major) contribution  to the
emission comes  from an inner  component is surprising.  Their massive
counterparts, the  Herbig~Be stars, show clear  evidence of optically
thick  gas  shielding  part  of  the stellar  emission  enabling  dust
sublimation at a closer distance from the star \citep{kraus08, bagnoli10}.  Our study indicates however that the inner region
in \hr~displays a discontinuity in surface
brightness that  is inconsistent with a  dense continuous accretion
disk extending up to the star. 

\indent The nature of the emission in the inner disk is unclear. A hot gaseous
disk as well  as refractory grains have been  previously suggested for
other      Herbig~Ae      stars     \citep{tannirkulam,      eisner09,
  benisty10}. Ascertaining the emission nature is beyond the scope of the paper, as even simple models could not
be accurately tested against  our interferometric measurements, because of the  complexity of the visibility curve.   However,  the  contribution  and
properties of a gaseous disk could be studied using high resolution NIR spectra.  A dense and
cold gaseous disk would produce a high level of NIR continuum emission
together with strong molecular lines, while non-LTE tenuous layers of 
atomic  or ionized  gas would  generate a  low level  of  continuum and
strong  emission lines.  An  accurate estimate  of the  mass accretion
rate based on an analysis across a broad range of wavelengths would also be useful in
constraining the properties of the inner disk.  On the other hand,
refractory dust  grains could also  be responsible for  this continuum
emission but they would be required to survive at temperatures well above their
tabulated   sublimation   temperatures.    Self-consistent  models   of
dust-free gaseous disks and multi-dust population
should   be   computed  to   fully   understand  the   interferometric
observations. \\
\indent While the derived flux contributions of the two main features in the
$K$-band image and their extents are robust, the structure
of the ring being less clear.  We  added in a crude way the inner disk to
the ring, with no self-consistent computation of the dusty disk structure at the
opacity transition. It is unclear what  its exact structure would be when the  stellar light is partially  shielded by the  inner disk, but
it seems likely that its morphology would deviate from the classical
puffed-up rim case. \\
\indent  We have searched  for clear  trends  in time  variability that  could
explain part of the scatter in the visibility data \citep{eiroa02, sitko08}, because the
timescales for changes  in the structure of the dusty  disk and in the
dynamics  of the  gas  - such  as  dust grain  destruction or  growth,
Keplerian rotation, accretion through bursts, instabilities - are well
within the two year range of our observations.  More generally, any
variation  in the  stellar and/or  NIR  excess flux  would affect  the
visibility measurements.  We found no such trend, and this interesting
issue will be investigated  further with more precise measurements, since
instrumental problems related to absolute calibrations may be 
responsible for some of the scatter. Gathering 
within a short timescale (e.g.,  a month) such a large interferometric
dataset and obtaining contemporaneous  NIR photometry would however be
necessary to study this problem.\\
\indent To understand whether the properties derived for \hr~are common among Herbig~Ae stars, and if they can be interpreted as a sign of disk evolution, a larger sample of stars should be studied with an extensive \uv coverage at different wavelengths.  
To perform accurate and unambiguous image reconstruction at the milli-arsecond resolution, highly precise  measurements are required,  as well  as very  long baselines ($\geq$140~m). Improvements are also expected on the image reconstruction codes, to enable e.g., monochromatic images in the continuum fitting the SED without any assumption about the color of the emission. 
In this context,  the NIR 4-beam instruments PIONIER/VLTI, MIRC/CHARA, and GRAVITY/VLTI are  expected to  achieve tremendous  progress in the coming years and to enable variability study on timescales of a few months.

\begin{acknowledgements}
  We thank the VLTI team at Paranal.   M.B.  acknowledges funding  from INAF  (grant ASI-INAF I/016/07/0). We thank the anonymous referee for useful comments that help improved the manuscript. 
\end{acknowledgements}

\bibliographystyle{aa}
\bibliography{v856scoamber}

\begin{appendix}
\section{Details of the observations}
\begin{table*}[h]
\centering
\caption{Log of the observations, including the baseline  name, length (P$_{\rm{L}}$), position angle (PA), as well as the average seeing ($\Theta$) and coherence time ($\tau$).} 
\label{tab:obs}
\begin{tabular}{cccccc|cccccc|cccccc}
Date &  Name & P$_{\rm{L}}$ & PA  & $\Theta$  & $\tau$ & Date  & Name &
P$_{\rm{L}}$ &  PA  & $\Theta$ & $\tau$  &  Date  &  Name &  P$_{\rm{L}}$  &
PA & $\Theta$  & $\tau$ \\ 
& & (m)& ($^\circ$) & ('')& (ms) & & & (m)& ($^\circ$)&('')
 & (ms) & & & (m)& ($^\circ$) & ('') & (ms)\\
\hline
25/02/08 & K0-A0 &  126 & 56 & 0.9  & 2.8 & 15/05/09& G1-H0&69  & 11 & 1.0 & 4.7 &14/04/10&
D0-I1& 79& 89 & 1.0 & 6.8\\
& G1-A0 & 84 & 103 & &  & & G1-D0 & 71& -42 & & & &I1-G1&46 & -148 & \\
& K0-G1 & 90 & 16 & & & & D0-H0& 63& 76 && & &D0-G1& 67& 125 & \\

05/04/09&D0-H0  &  64   &  65  & 0.8 & 3.1 &  16/03/10&  G1-H0&   69&  -8  & 0.4 & 16.8 &
07/06/10&H0-I1 &40 &160 & 1.3 & 1.6\\
&G1-H0 & 69& 3 &  && & G1-D0& 60& -63 & & &  & I1-E0&68 & -58 & \\
&G1-D0 & 69 & -52 & &  & & D0-H0& 60 & 46 & & & & H0-E0&44 & -93 & \\

08/04/09& G1-A0  &88 & 109 & 1.2 & 1.8 &  17/03/10& G1-H0& 69& -2  & 0.5 & 18.4 & 11/06/10&
D0-H0& 63 & 71 & 0.7 & 8.0\\
&K0-A0 & 128 & 65 & &  & & G1-D0& 65 & -58 & & &  & D0-I1& 82 & 99 & \\
& K0-G1& 90 & 21 & & & & D0-H0& 63 & 56 & & &  & I1-H0& 40 & -32 & \\

21/04/09& H0-E0&  45& 85& 0.9 & 4.2 & 21/03/10& H0-G0&  31& -128 & 0.6 & 9.7 & 15/06/10&
K0-I1& 46 & -4 & 0.6 & 6.7\\
&G0-E0 & 15& 85 & & &  &H0-E0& 46& -128 & & &  & K0-G1 & 90 & 9 & \\
&H0-G0 & 30& 85& & &  & G0-E0& 15 & -128  & & & & I1-G1 & 46 & 20 &\\

30/04/09& G1-H0& 69& -10 & 0.9 & 4.4 &  10/04/10& H0-E0& 43 & -147 & 0.6 & 22.9 \\
 & G1-D0 & 58 & -65 & &  & & I1-E0& 50 & -108 & \\
&D0-H0 & 59 & 43 & & &  & H0-I1& 32& 130 & \\

\end{tabular}
\end{table*}

\begin{figure*}[h]
 \centering
\begin{tabular}{cc}
 \includegraphics[width=0.35\textwidth]{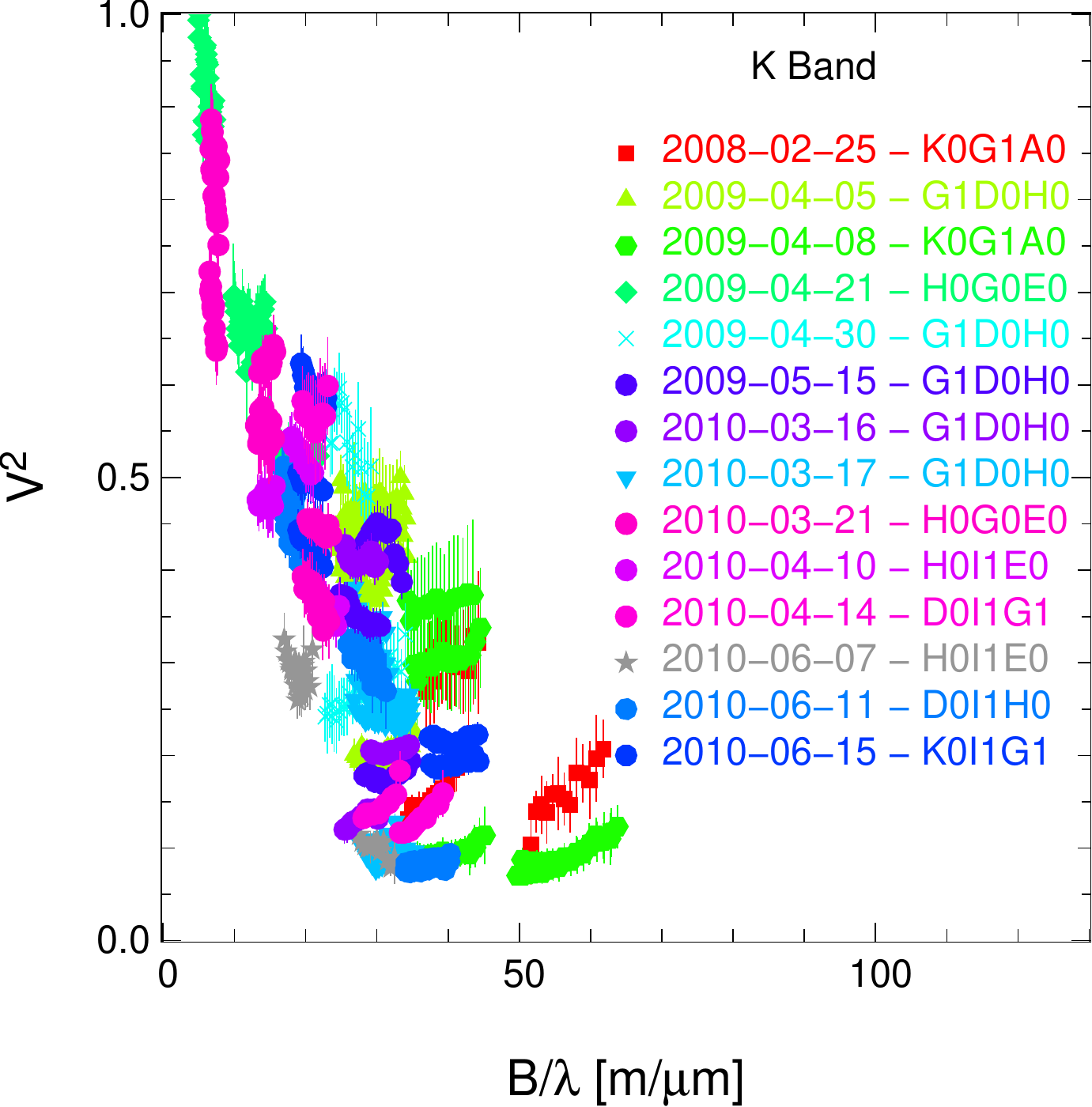}
 &
 \includegraphics[width=0.35\textwidth]{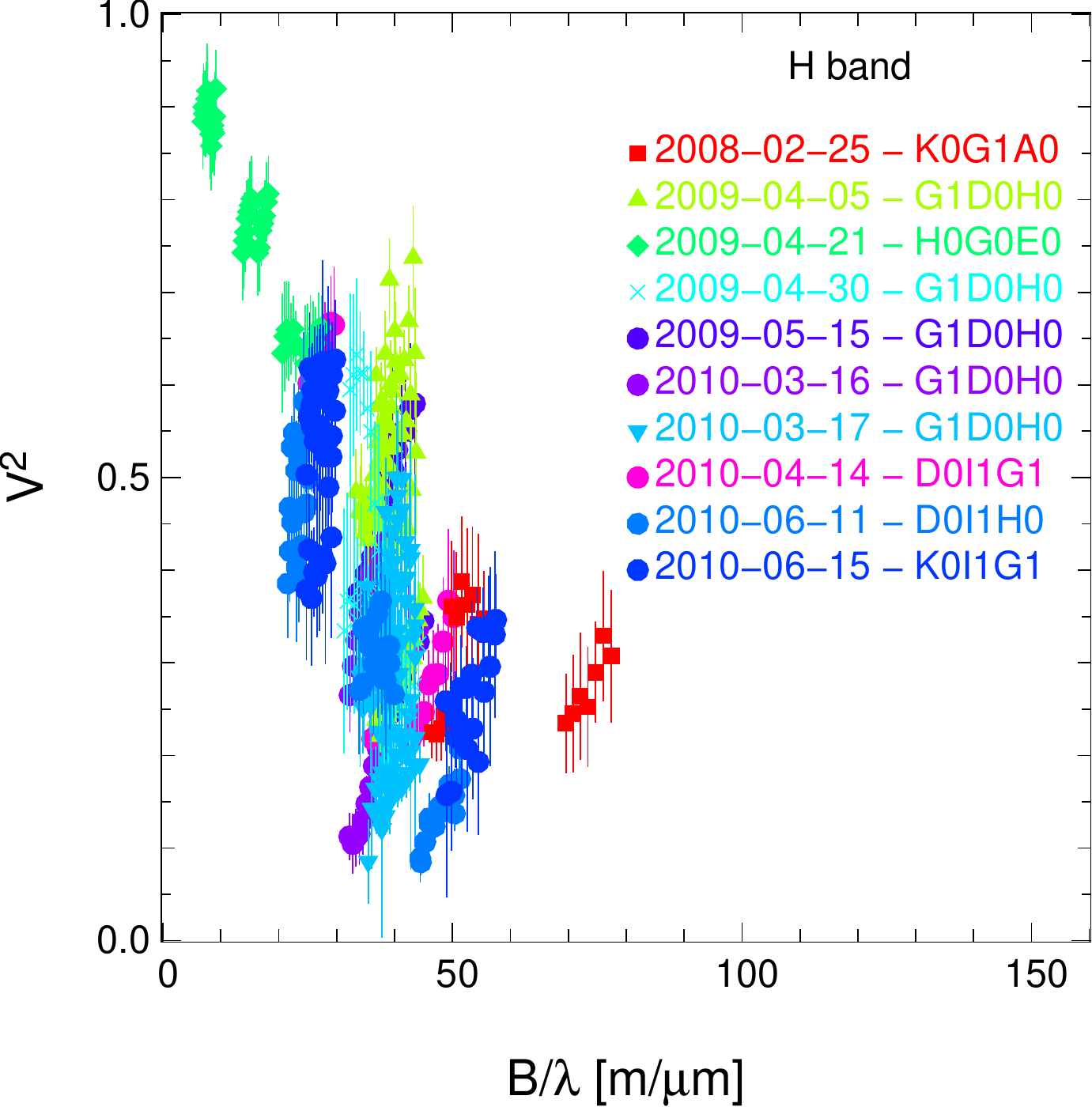}
\end{tabular}
  \caption{\label{fig:V2obs}  Observed  spectral  visibilities  against
 spatial frequencies measured in the $K$  band (left) and
 the  $H$  band  (right).   Each  night/dataset is  indicated  with  a
 specific color.  } 
\end{figure*}

\begin{figure*}[h]
 \centering
\begin{tabular}{cc}
 \includegraphics[width=0.35\textwidth]{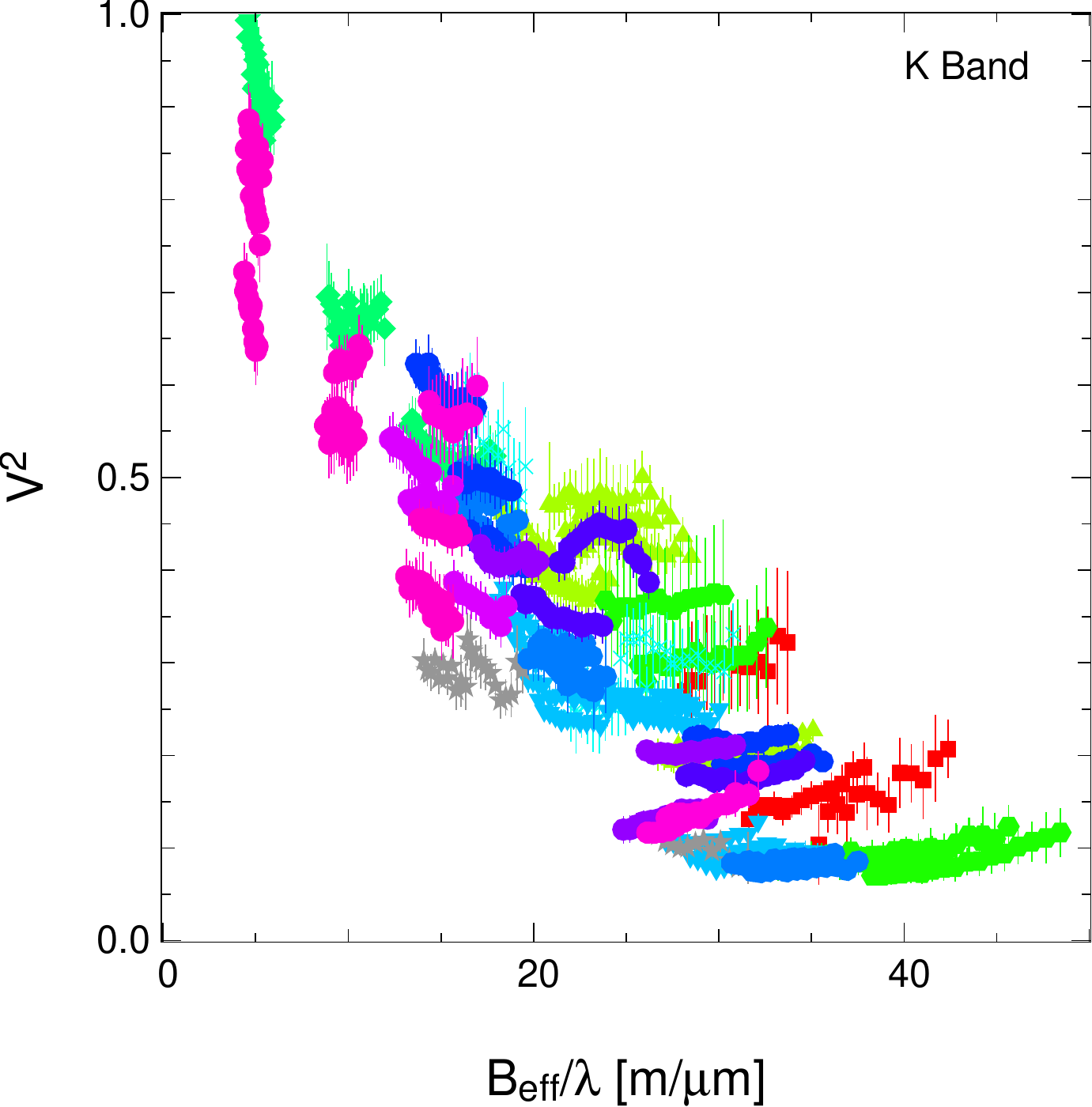}
 &
 \includegraphics[width=0.35\textwidth]{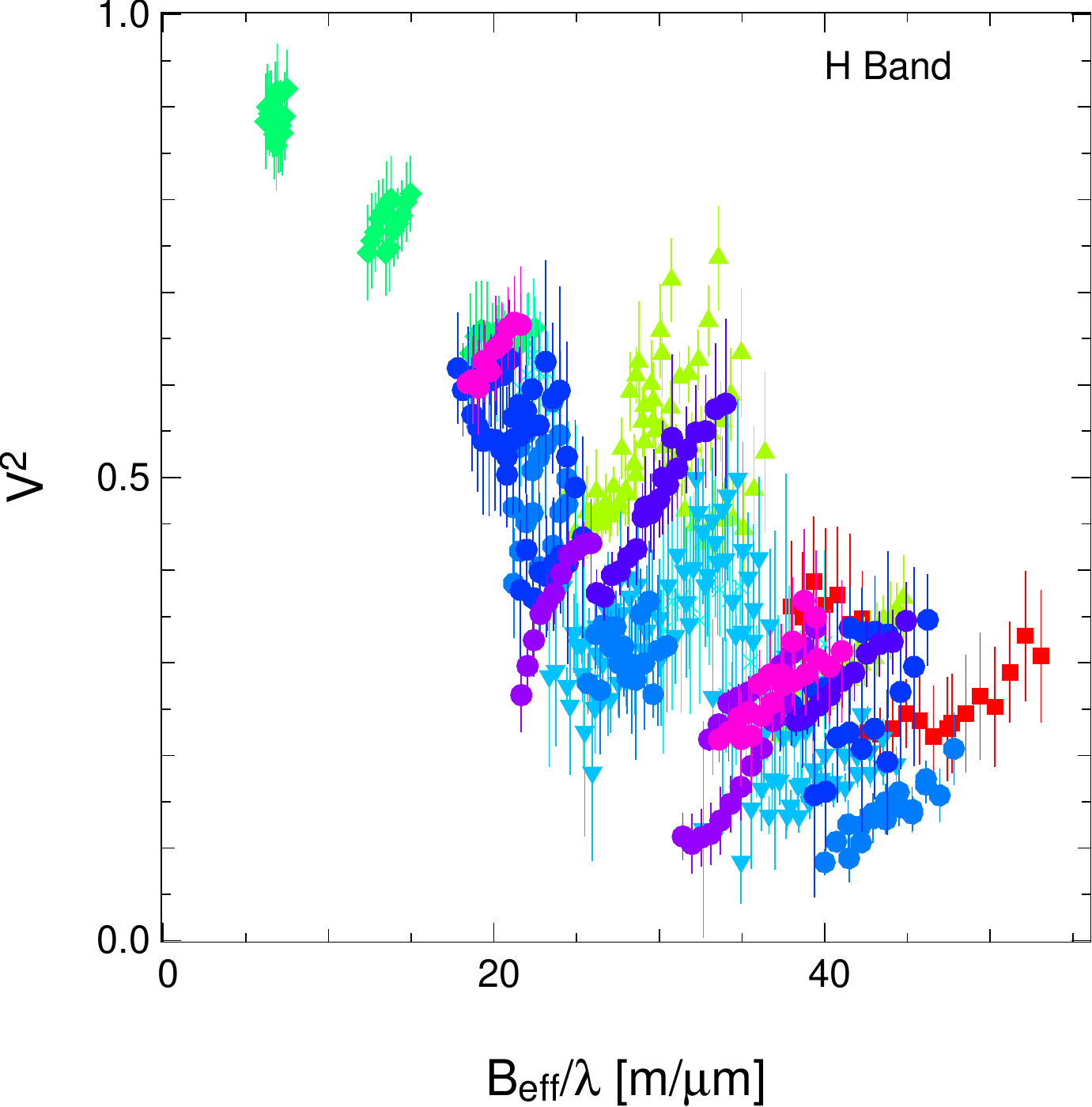}
\end{tabular}
  \caption{\label{fig:V2obs2}  Observed  spectral  visibilities  against
 effective spatial frequencies measured in the $K$  band (left) and
 the  $H$  band  (right) and assuming i=48$^\circ$ and PA=135$^\circ$.  Each  night/dataset is  indicated  with  a
 specific color.  } 
\end{figure*}

\begin{figure*}[!t]
 \centering
\begin{tabular}{ccc}
 \includegraphics[width=0.32\textwidth]{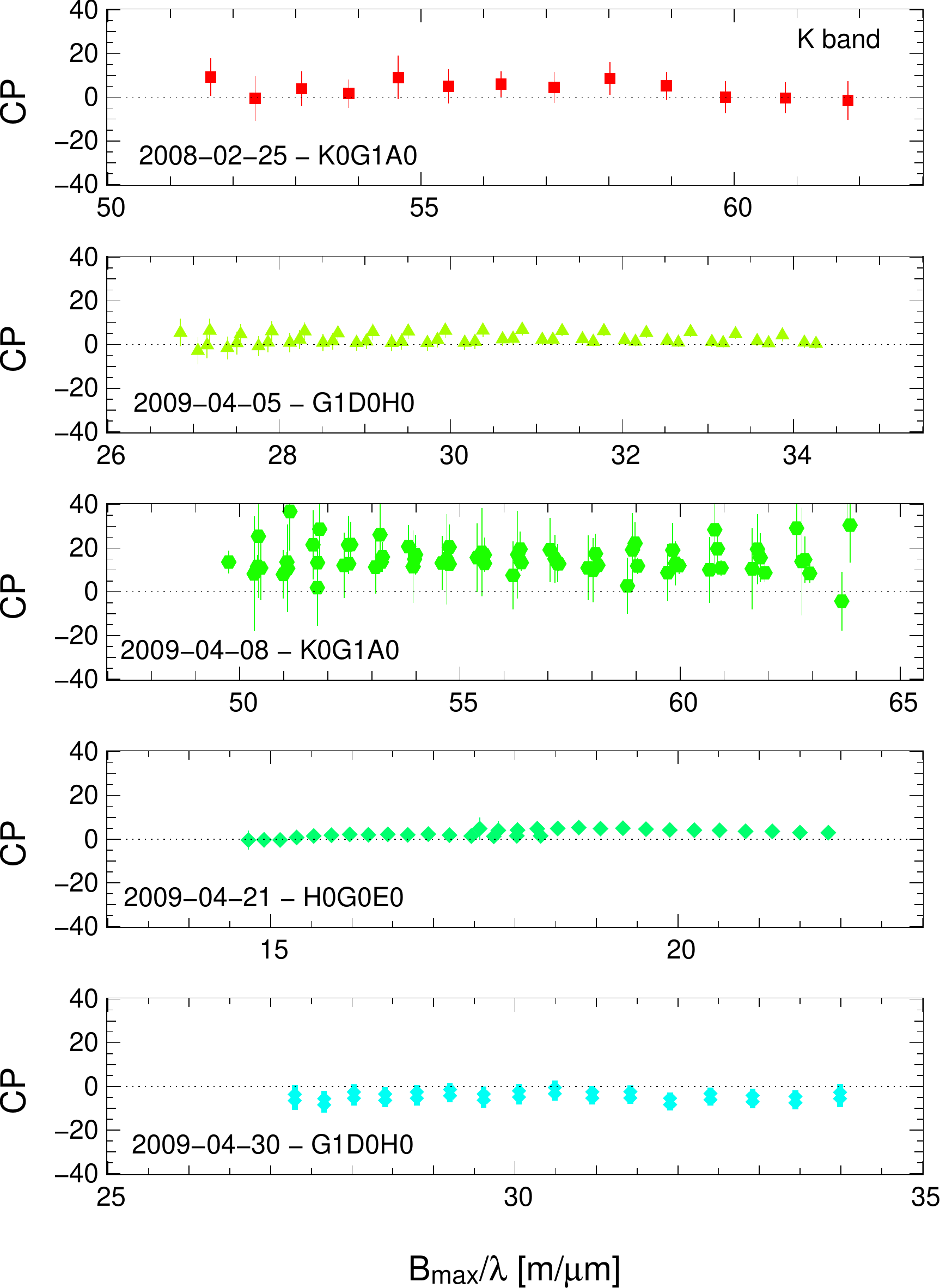}
 &
 \includegraphics[width=0.315\textwidth]{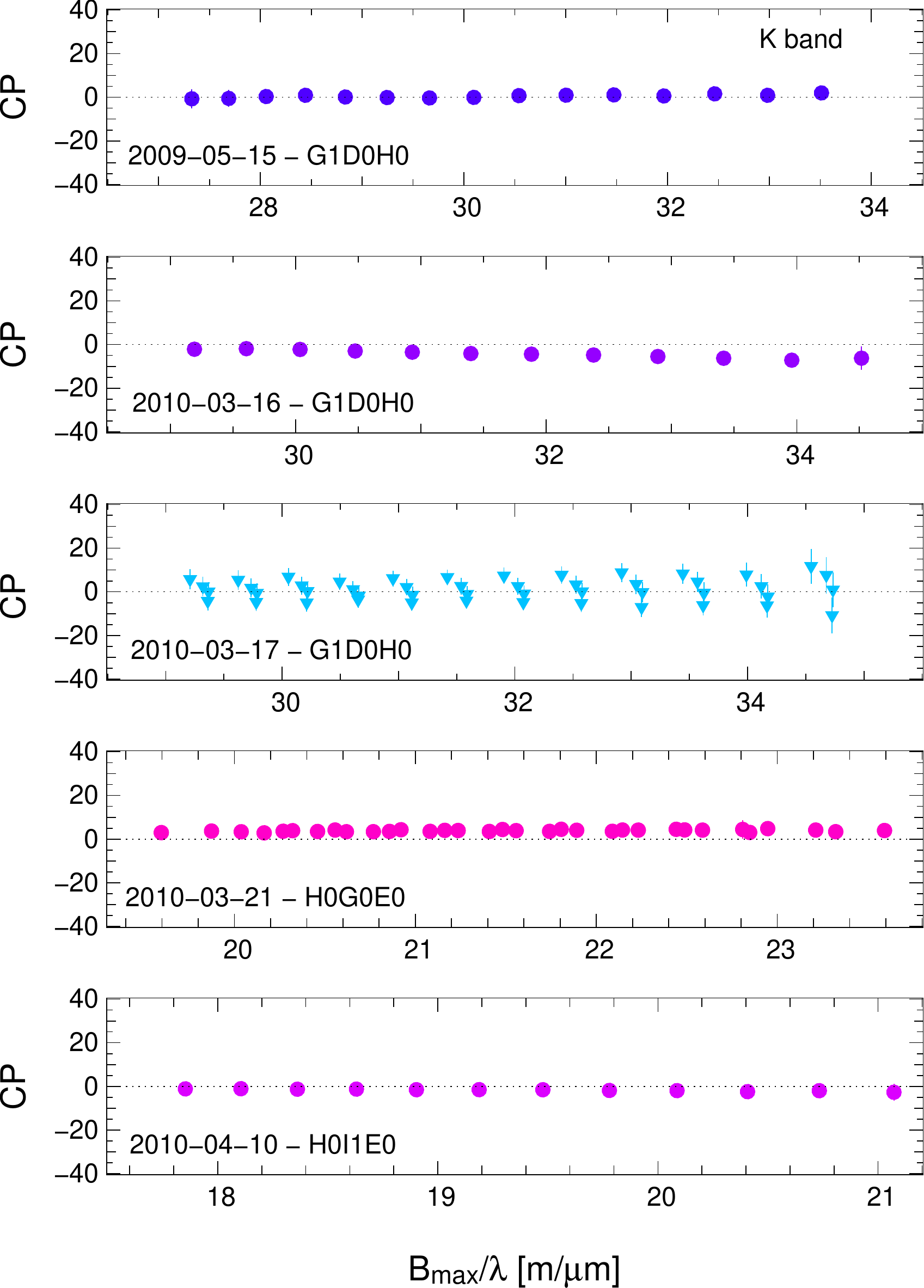}
&
 \includegraphics[width=0.32\textwidth]{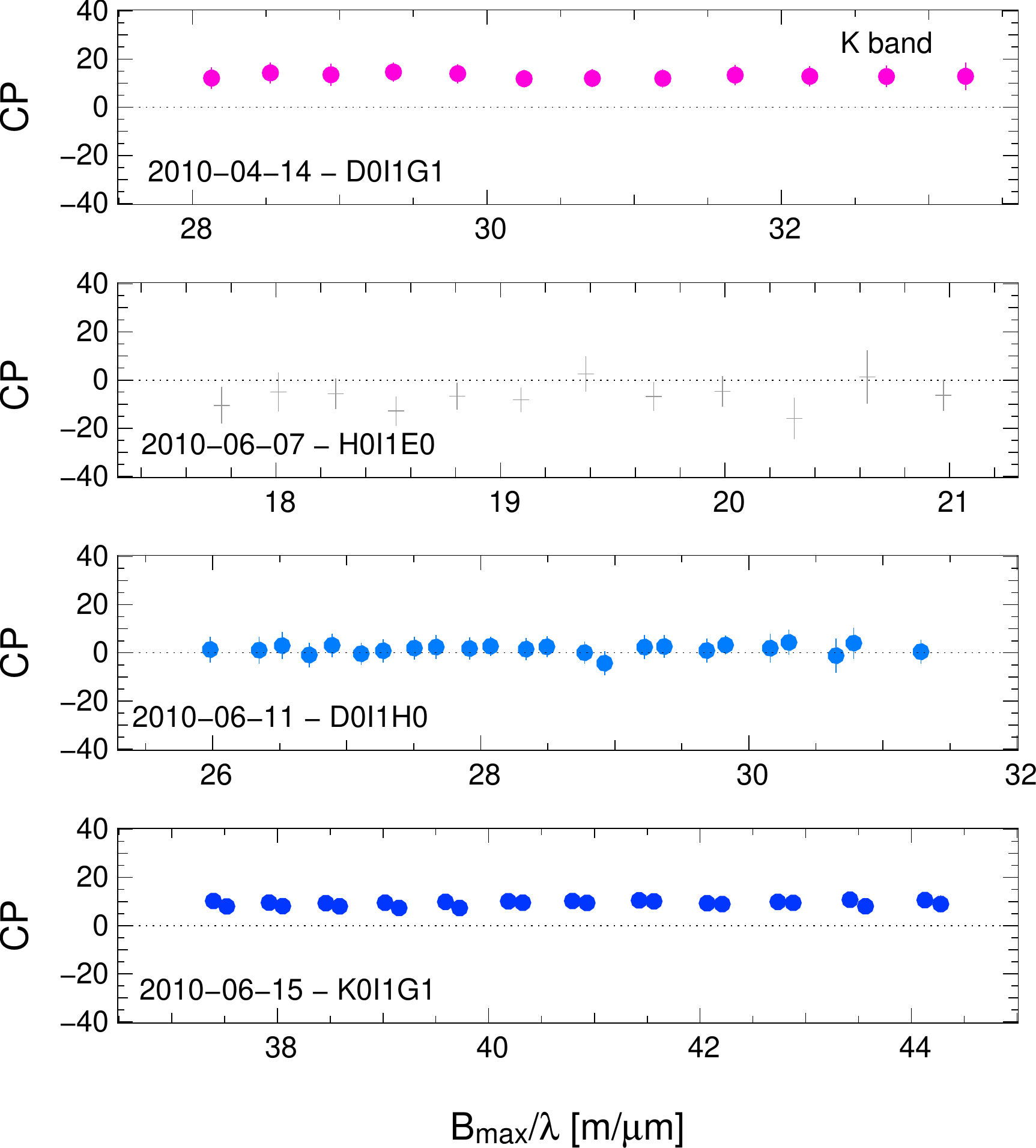}
\\
 \includegraphics[width=0.32\textwidth]{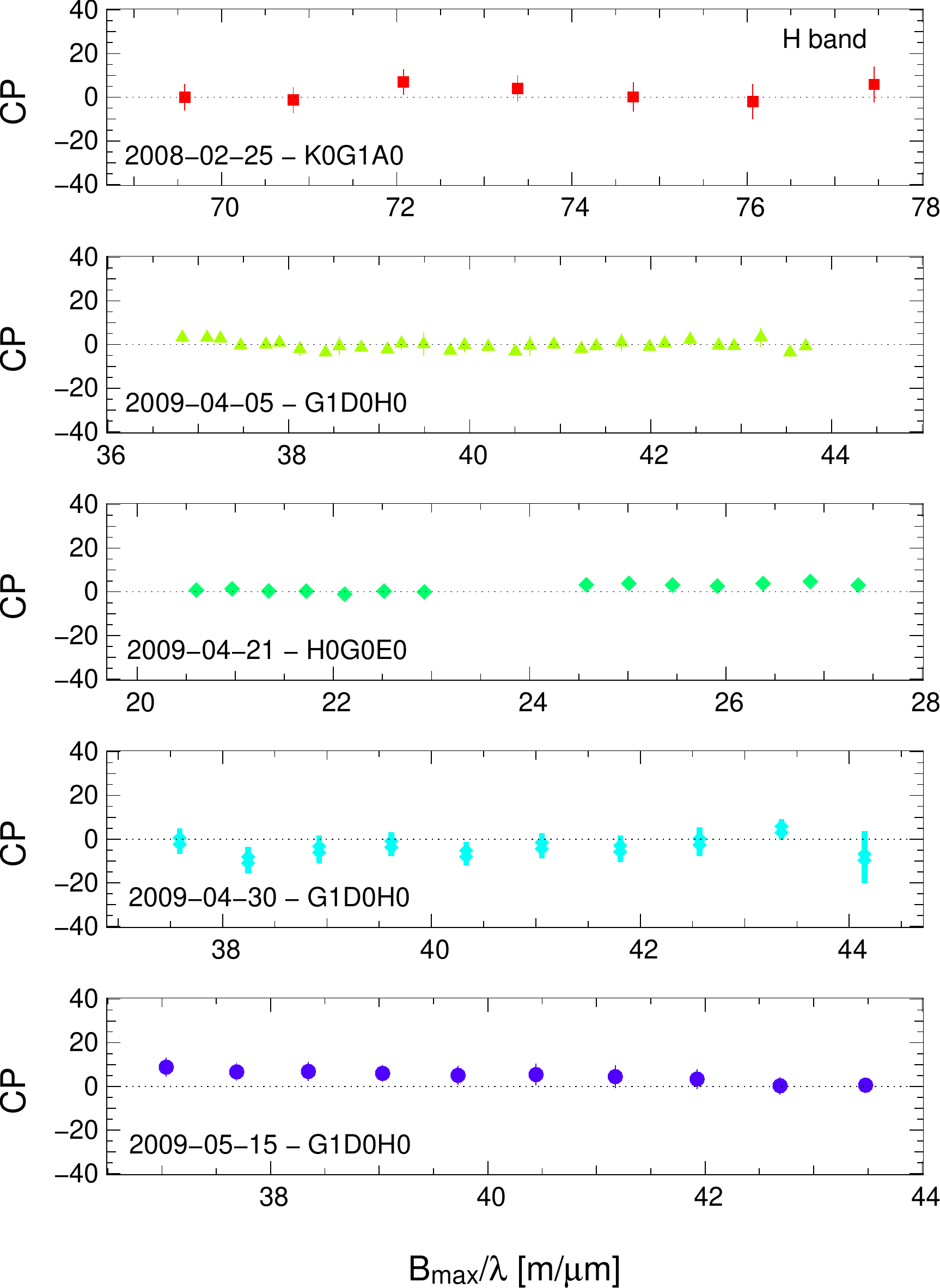}
 &
 \includegraphics[width=0.32\textwidth]{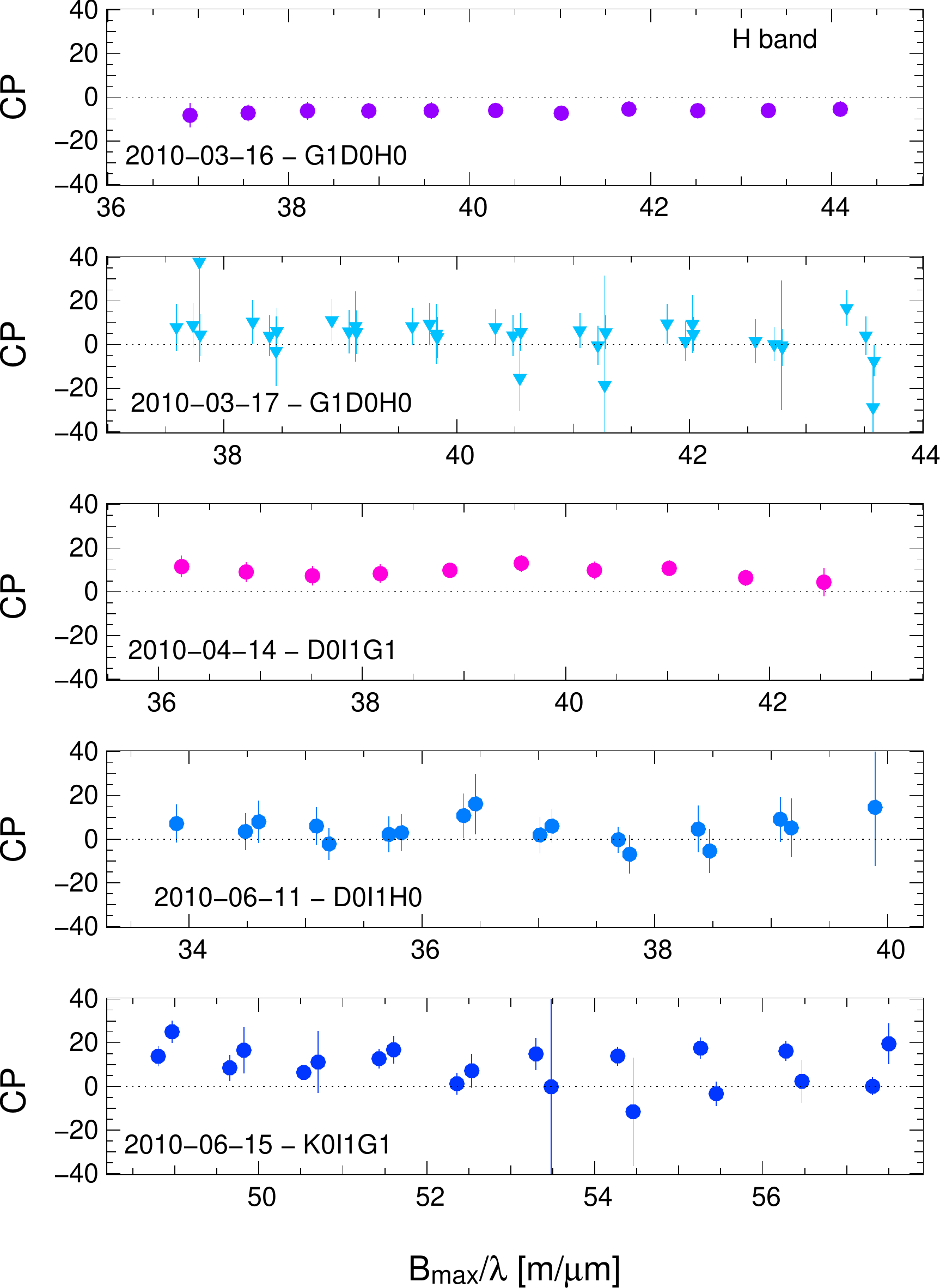}
\end{tabular}
  \caption{\label{fig:CPobs}  Observed  spectral  closure  phases  (in
 degrees) against
maximum spatial frequencies measured in the $K$ band (upper panels) and the $H$ band
 (lower  panels).  Each  night/dataset  is indicated  with a  specific
 color, according to the same color code as in Fig.~\ref{fig:V2obs}.} 
\end{figure*}

\begin{figure*}[h]
 \centering
%\begin{tabular}{cc}
 \includegraphics[width=0.35\textwidth]{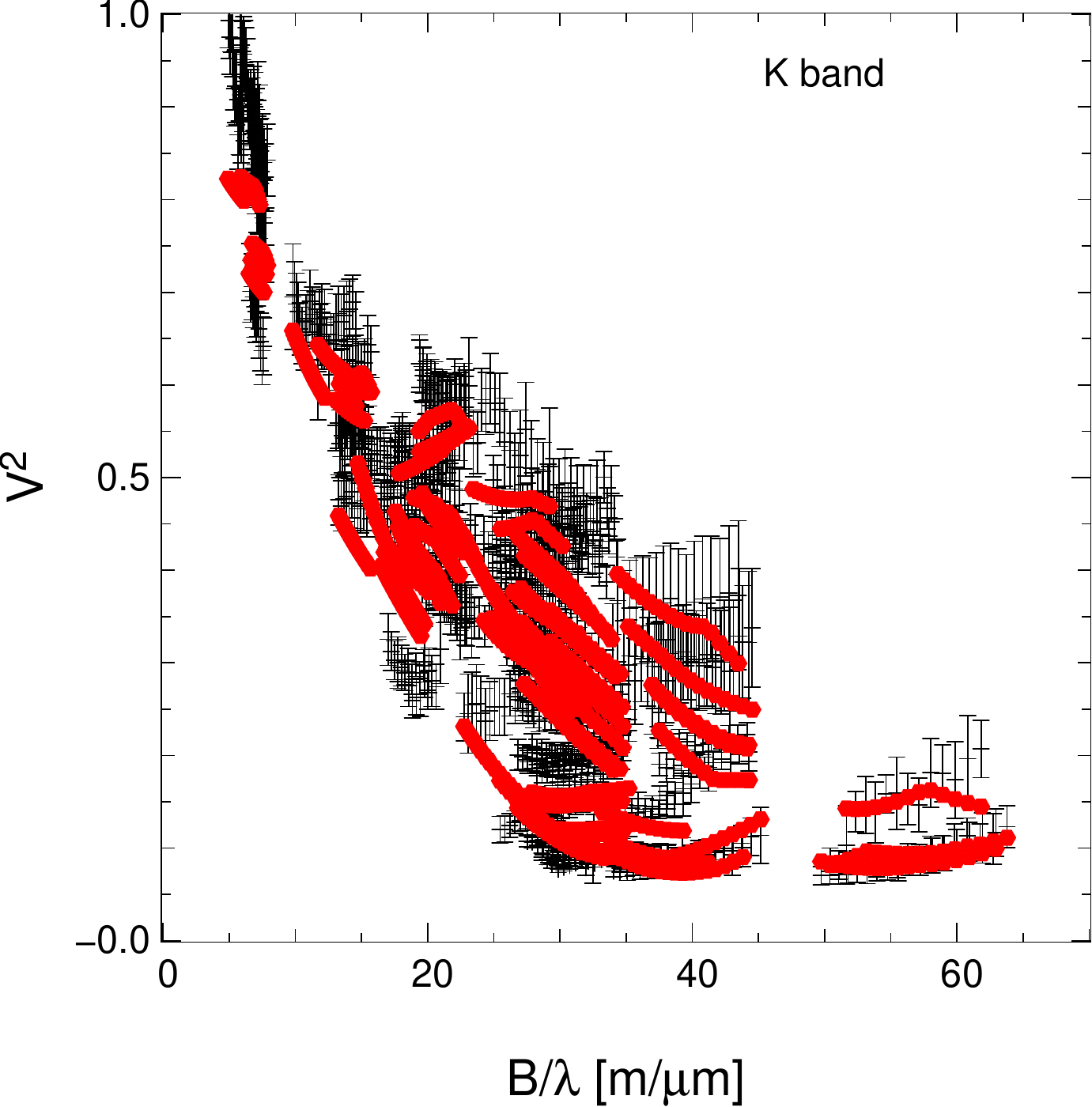}
 %&
 %\includegraphics[width=0.31\textwidth]{HR5999_fitMiRA_vis2_H.pdf}
%\end{tabular}
  \caption{\label{fig:V2mira} Squared visibilities measured (black) and computed from the reconstructed image (red filled circles) for the K~band. } 
\end{figure*}

%\begin{figure*}[h]
% \centering
%\begin{tabular}{cc}
% \includegraphics[width=0.35\textwidth]{HR5999_fitMiRA_cp_K.pdf}
% &
% \includegraphics[width=0.35\textwidth]{HR5999_fitMiRA_cp_H.pdf}
%\end{tabular}
%  \caption{\label{fig:CPmira}   \textbf{Closure phases measured (black) and computed from the reconstructed image (red filled circles) for the K~band (left) and H~band (right). }} 
%\end{figure*}

\begin{figure*}[!t]
 \centering
\begin{tabular}{ccc}
 \includegraphics[width=0.32\textwidth]{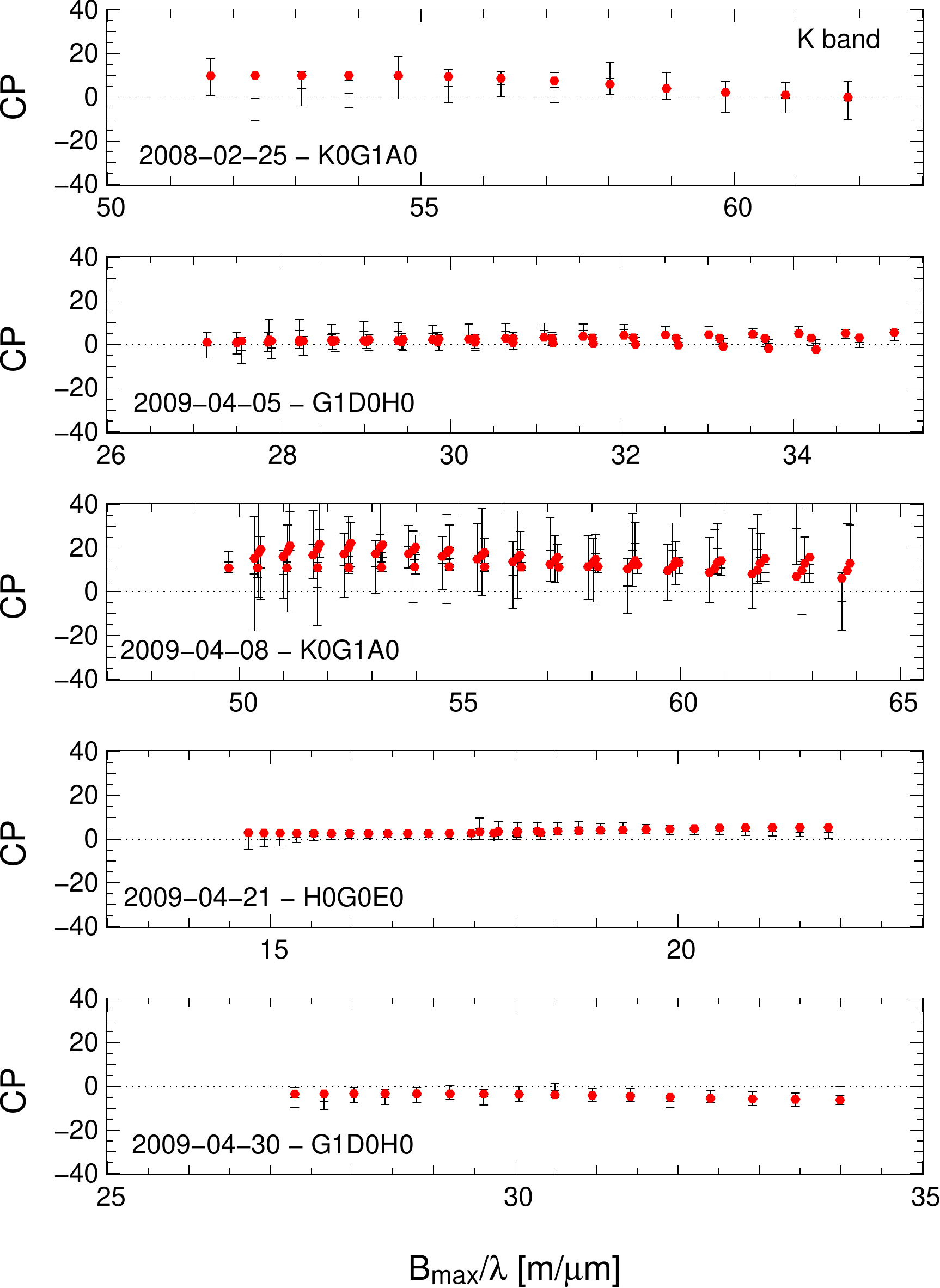}
 &
 \includegraphics[width=0.315\textwidth]{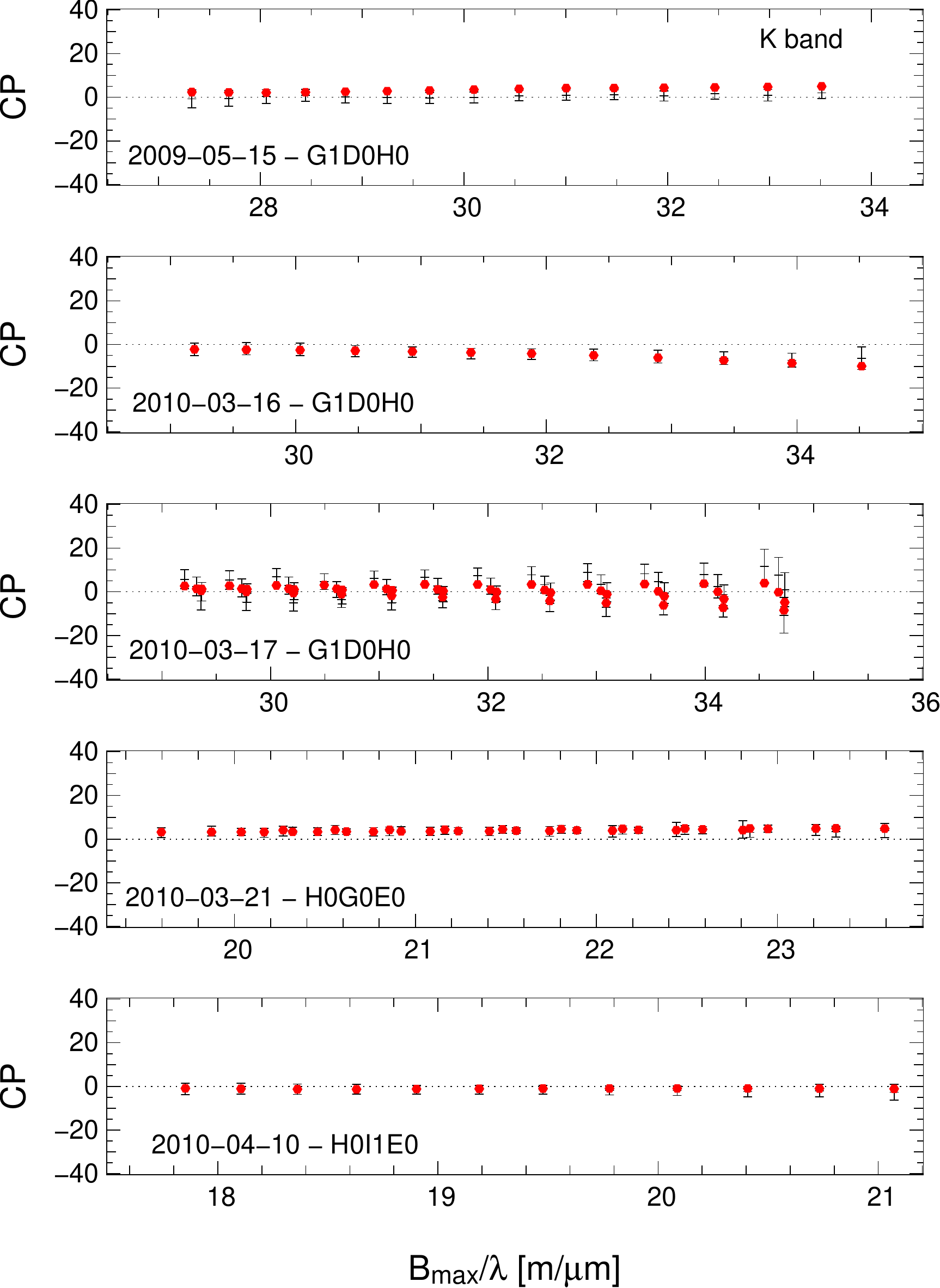}
&
 \includegraphics[width=0.32\textwidth]{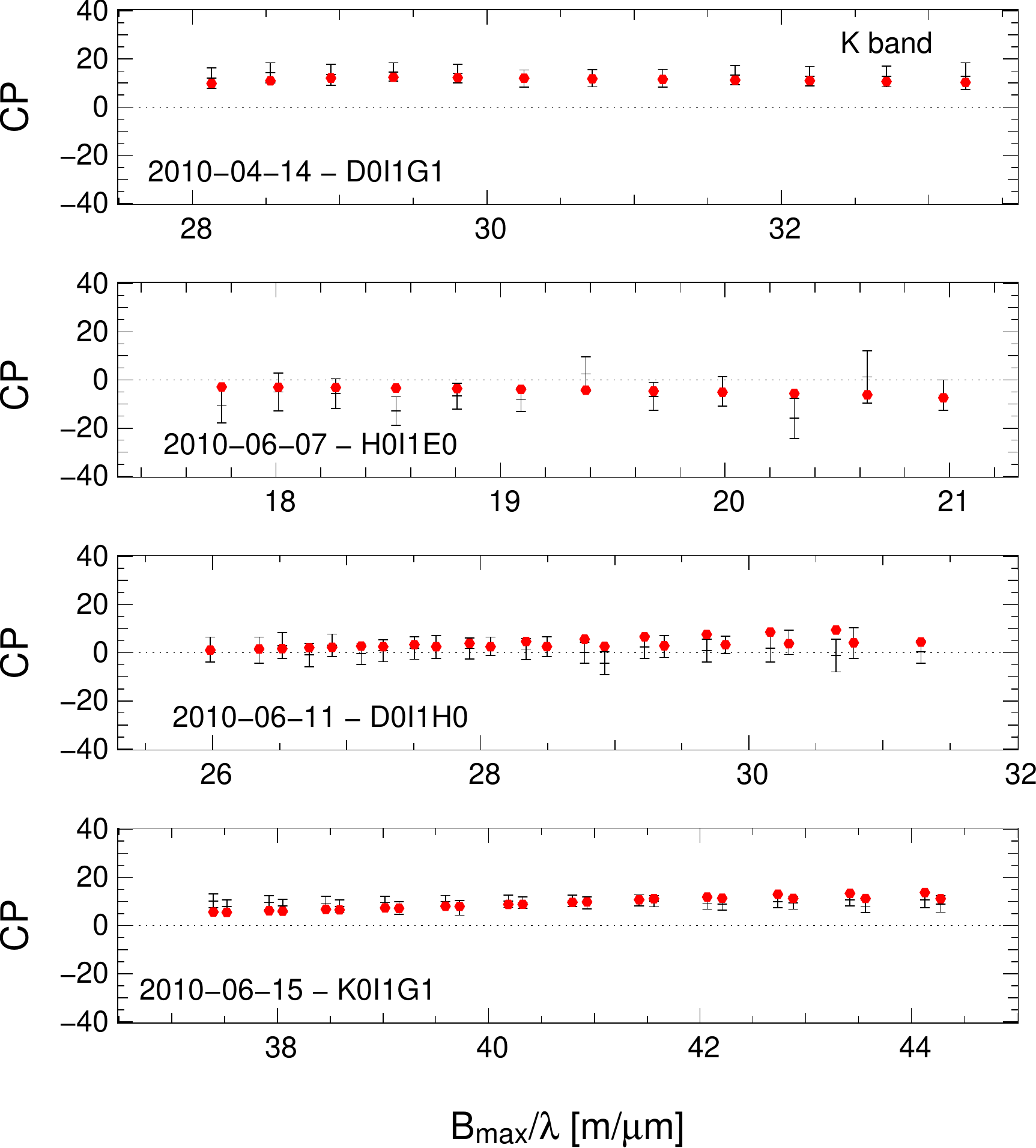}
%\\
% \includegraphics[width=0.32\textwidth]{HR5999_fitMiRA_CP_H_panel1}
% &
% \includegraphics[width=0.32\textwidth]{HR5999_fitMiRA_CP_H_panel2}
\end{tabular}
  \caption{\label{fig:CPmira} Closure phases measured (black) and computed from the reconstructed image (red filled circles). }
\end{figure*}

\end{appendix}

\end{document}